 \documentclass[aip,pop,reprint]{revtex4-1}
 \usepackage{dcolumn}
 \usepackage{bm}
 \usepackage{amsmath}
\newcommand{\mx}{\textrm{\scriptsize max}}

\newcommand{\bb}{\textrm{\scriptsize B}}

\newcommand{\vc}{\mathbf}
\usepackage{graphicx}
\usepackage{color}

\begin{document}

% \preprint{UW-CPTC 09-5}

\title{dc electrical conductivity in strongly magnetized plasmas}


\author{Scott D.\ Baalrud}
%\author{J\'{e}r\^{o}me Daligault$^2$}
%\homepage{http://http://www.lanl.gov/projects/dense-plasma-theory}
\email{baalrud@umich.edu}
\affiliation{Department of Nuclear Engineering and Radiological Sciences, University of Michigan, Ann Arbor, MI 48109, USA}
%\affiliation{Department of Physics and Astronomy, University of Iowa, Iowa City, Iowa 52242, USA}

\author{Trevor Lafleur} 
\affiliation{PlasmaPotential-Physics Consulting and Research, Canberra, ACT 2601, Australia} 
%\affiliation{$^2$PlasmaPotential}


\date{\today}

\begin{abstract}

A generalized Ohm's law is derived to treat strongly magnetized plasmas in which the electron gyrofrequency significantly exceeds the electron plasma frequency. 
The frictional drag due to Coulomb collisions between electrons and ions is found to shift, producing an additional transverse resistivity term in the generalized Ohm's law that is perpendicular to both the current ($\vc{J}$) and the Hall ($\vc{J} \times \vc{B}$) direction. 
In the limit of very strong magnetization, the parallel resistivity is found to increase by a factor of 3/2, and the perpendicular resistivity to scale as $\ln (\omega_{ce} \tau_e)$, where $\omega_{ce} \tau_e$ is the Hall parameter. 
Correspondingly, the parallel conductivity coefficient is reduced by a factor of 2/3, and the perpendicular conductivity scales as $\ln(\omega_{ce} \tau_e)/(\omega_{ce} \tau_e)^2$. 
These results suggest that strong magnetization significantly changes the magnetohydrodynamic evolution of a plasma. 

\end{abstract}

%\pacs{52.25.Fi,52.27.Gr,52.65.Yy}

% 52.25.Fi Transport properties
% 52.27.Gr Strongly coupled plasmas 
% 52.65.Yy Molecular dynamics methods

%\keywords{}

\maketitle


\section{Introduction}

Electrical conductivity describes how electric currents are driven in response to weak electric fields in a material. 
It is especially complicated in magnetized plasmas because the conductivity becomes a tensor with coefficients that depend on the local magnetic field, and the Hall effect causes the currents to be misaligned with the electric field~\cite{BraginskiiRPP1965}. 
Despite these complications, it is important to accurately model because the generation of electrical currents is central to many of the most important research questions in plasma physics, including magnetic field generation~\cite{Kulsrud_2008,FoxPOP2018}, magnetic field evolution~\cite{gurnett_bhattacharjee_2005}, energy dissipation~\cite{RaxPOP2019}, magnetic reconnection~\cite{YamadaRMP2010}, dynamo amplification~\cite{plunian_alboussiere_2021}, and magnetic confinement~\cite{Wesson:1427009}. 

Current understanding of electrical conduction in collisional regimes is largely based on the Chapman-Enskog~\cite{Chapman:1991} solution of the plasma kinetic equation~\cite{LandauPZS1936,Spitzer:1953}, as was summarized by Braginskii~\cite{BraginskiiRPP1965}. 
This leads to an expression for Ohm's law
\begin{equation}
\label{eq:ohm}
\vc{J} = \vc{\sigma} \cdot \vc{E}^\prime = \sigma_\parallel \vc{E}_\parallel^\prime + \sigma_\perp \vc{E}_\perp^\prime + \sigma_\wedge \vc{E}_\wedge^\prime
%J_\alpha = \sigma_{\alpha \beta} E_\beta
\end{equation} 
that is a linear relationship between the current density $\vc{J}$ and the electric field in the rest frame of the fluid $\vc{E}^\prime = \vc{E} + \vc{V} \times \vc{B}$. The conductivity tensor $\vc{\sigma}$ depends on three independent coefficients $\sigma_\parallel$, $\sigma_\perp$ and $\sigma_\wedge$, which the theory computes with explicit closed form expressions that depend on the plasma density, temperature, and magnetic field strength.~\cite{NRL1983}  
%It is a linear relationship because the coefficients are independent of $\vc{J}$ and $\vc{E}$. 

The plasma kinetic theories that this traditional transport model is based on (Boltzmann~\cite{Ferziger1972}, Landau~\cite{LandauPZS1936} or Lenard-Balescu~\cite{LenardAP1960,BalescuPOP1960} equations) are derived from an assumption that the plasma is weakly magnetized in the sense that the gyrofrequency is much smaller than the plasma frequency for each species in the plasma: $\omega_c \ll \omega_p$, where $\omega_c = qB/m$ and $\omega_p = \sqrt{q^2 n/ \epsilon_o m}$. 
In this limit, gyromotion is negligible at the microscopic scales at which collisions occur (within a Debye sphere) since $r_c \gg \lambda_D$ where $r_c = v_T/\omega_c$ is the thermal gyroradius, $\lambda_D = v_T/(\sqrt{2} \omega_p)$ is the Debye length, and $v_T = \sqrt{2k_BT/m}$ is the thermal speed. 
As a consequence, the magnetic field does not influence the collision operator in the traditional kinetic theory. 
Unequal parallel ($\sigma_\parallel$) and perpendicular ($\sigma_\perp$) conductivity coefficients arise from the convective term in the kinetic equation, as the velocity distribution function of particles (particularly electrons) is more easily distorted along the magnetic field than across it.~\cite{Ferziger1972} 
This term also gives rise to the Hall effect responsible for the off-diagonal components $\sigma_\wedge$. 

Here, we consider strongly magnetized plasmas in which the electron gyrofrequency significantly exceeds the electron plasma frequency: $\beta_e \gg 1$, where
\begin{equation}
\beta_e \equiv \frac{\omega_{ce}}{\omega_{pe}} .
\end{equation} 
Electron gyromotion is sub-Debye scale at these conditions, indicating that electrons are magnetized at the microscopic scale at which collisions occur. 
Traditional plasma kinetic theories do not apply in this circumstance. 
Generalizations have been proposed to treat strong magnetization, including linear response (Lenard-Balescu-type) approaches,~\cite{RostokerPF1960,NersisyanPRE2003,Nersisyan:2007} collision (Boltzmann-type) approaches~\cite{OneilPF1983,NersisyanPRE2009,JosePOP2020}, and phase-space convection-diffusion (Fokker-Planck type) approaches~\cite{MontgomeryPF1974,WarePRL1989,CohenPOP2019,DubinPOP2014}. 
A comprehensive calculation of electrical conductivity would require solving one of these generalized kinetic equations using a Chapman-Enskog~\cite{Chapman:1991} or Grad-type~\cite{Grad1958} perturbative method. 
This is a formidable challenge due to the increased complexity of the generalized collision operators, and has not yet been achieved. 
Here, we focus on the electron-ion collision contribution to the conductivity and apply a moment approach that corresponds to the first-order of Grad's method. 
This focuses attention on the fact that the Coulomb collision frequency comes to depend on the magnetic field strength and orientation in a strongly magnetized plasma. 
It also allows a direct connection between the electrical conductivity and the friction force on a test ion. 

Recent work has considered in detail the friction force on a test ion moving through a strongly magnetized electron background, revealing novel behaviors~\cite{LafleurPPCF2019,LafleurPPCF2020,BernsteinPRE2020,JosePOP2020,Nersisyan:2007}. 
Friction is normally expected to act antiparallel to the velocity of the test charge;~\cite{CerecedaPOP2005,Nersisyan:2007} an expectation set by predictions of kinetic theory in the weakly magnetized regime. 
In contrast to this expectation, linear response theory was used to predict that strong magnetization of the background plasma causes the friction force to shift, obtaining a transverse component that is perpendicular to the velocity of the test charge in the plane formed by the velocity and magnetic field~\cite{LafleurPPCF2019}. 
This prediction was later tested using first-principles molecular dynamics simulations, which were found to agree well with the theoretical predictions~\cite{BernsteinPRE2020}. 
It causes non-intuitive effects on the motion of a test charge, such as causing the gyroradius of a fast particle to increase in time~\cite{LafleurPPCF2019}. 

Applying these results, we show that these new features of single particle motion translate to changing Ohm's law. 
A generalization in terms of resistivity shows that the transverse force leads to a transverse resistivity coefficient. 
Strong magnetization is found to cause the perpendicular resistivity coefficient to contain a logarithmic nonlinearly whereby it depends on the perpendicular current density $J_\perp$. 
However, the nonlinearity is a weak logarithmic function, and in the limit that the Hall parameter ($\omega_{ce} \tau_e$) is sufficiently large it becomes negligible. 
In this limit, and expressing Ohm's law in a coordinate system referenced to the direction of the magnetic field, the resistivity coefficients are found to be
\begin{subequations} 
\label{eq:eta_sum}
\begin{align}
\eta^\textrm{s}_\parallel &\approx \frac{3}{2} \eta^\textrm{w}_\parallel , \\ 
\eta^\textrm{s}_\perp &\approx \frac{3}{2} \ln (\omega_{ce} \tau_e) \eta^\textrm{w}_\perp, \\
\eta^\textrm{s}_\wedge &\approx \eta^\textrm{w}_\wedge . 
\end{align}
\end{subequations} 
where $\eta^\textrm{s}$ are the resistivity coefficients in the strongly magnetized limit, $\eta^\textrm{w}$ are the traditional resistivity coefficients computed from the first order of the Grad (or Chapman-Enskog) solution of the plasma kinetic equation (referred to here as the weakly magnetized regime), and $\omega_{ce} \tau_e$ is the Hall parameter, where 
\begin{equation}
\label{eq:tau_e}
\tau_e = \frac{3}{4\sqrt{2\pi}} \frac{(4\pi \epsilon_o )^2 \sqrt{m_e} (k_BT)^{3/2}}{n e^4 \ln \Lambda} 
\end{equation}
is the electron Coulomb collision time, $\ln \Lambda = \ln (\Gamma_e^{-3/2}/\sqrt{3})$ is the Coulomb logarithm, 
\begin{equation}
\Gamma_e = \frac{e^2/a_e}{ 4\pi \epsilon_o k_BT} 
\end{equation}
is the electron Coulomb coupling parameter, and $a_e = (3/4\pi n_e)^{1/3}$ is the average distance between electrons. 
The results of Eq.~(\ref{eq:eta_sum}) suggest that the magnitude of the perpendicular resistivity coefficient is significantly larger in the strongly magnetized limit. 

By inverting the resistive form of Ohm's law, conductivity coefficients of the form in Eq.~(\ref{eq:ohm}) are obtained. 
In the limit of a large Hall parameter, these are found to be 
\begin{subequations}
\label{eq:sigma_sum}
\begin{align}
\sigma^\textrm{s}_\parallel &\approx \frac{2}{3} \sigma^\textrm{w}_\parallel , \\
\sigma^\textrm{s}_\perp &\approx \frac{3}{2} \ln (\omega_{ce} \tau_e) \sigma^\textrm{w}_\perp , \\
\sigma^\textrm{s}_\wedge &\approx \sigma^\textrm{w}_\wedge . 
\end{align}
\end{subequations}  
This result suggests that strong magnetization enhances the current produced in the direction perpendicular to the magnetic field. 
In other words, magnetization at the scale of collisions promotes cross field electron transport. 

The linear response approach that these results are based on applies to plasmas that are weakly coupled $\Gamma_e \ll 1$. 
It also is expected to require that the gyroradius be larger than the distance of closest approach in a binary collision: $r_c > r_L = \sqrt{2} e^2/(4\pi \epsilon_o k_BT)$, which implies $\beta_e \lesssim 1/(\sqrt{6} \Gamma^{3/2})$. 
A recent generalization of the Boltzmann equation to account for strong magnetization does not have these restrictions, but is more computationally intensive to evaluate~\cite{JosePOP2020}. 
Future developments will be needed to apply this approach to address stronger magnetization and coupling regimes. 
An identification and explination for the different regimes in coupling-magnetization parameter space is provided in Ref.~\onlinecite{BaalrudPRE2017}. 

Although strong magnetization is not the norm, there are many important examples of laboratory and astrophysical plasmas in which it is encountered. 
A prominent laboratory example is non-neutral plasmas~\cite{DubinPOP1998}, such as the plasmas in antimatter traps~\cite{FajansPOP2020}, which can reach very strongly magnetized conditions. 
A host of other experiments may reach strongly magnetized conditions in the near future, including ultracold neutral plasmas~\cite{ZhangPRL2008,TiwariPOP2018,GormanPRL2021,GuthrieA2021}, dusty plasmas~\cite{ThomasPPCF2012,BonitzPSST2012,HartmannPRE2019}, plasmas in highly compressed magnetized inertial confinement fusion experiments~\cite{GomezPRL2014,GomezPRL2020}, and in intense laser-matter interaction experiments~\cite{TatarakisPOP2002}. 
Regimes in which electrons reach a marginal level of strong magnetization ($\beta_e \sim 1$) are more common, including electrons in tokamak experiments~\cite{AymarPPCF2002}. 
Strong magnetization of varying degrees can be reached in astrophysical plasmas, including planetary magnetospheres~\cite{2004jpsm.book..593K}, and the atmospheres of white dwarf~\cite{ValyavinN2014} and neutron stars~\cite{HardingRPP2006,PotekhinSSR2015}. 
A summary of example parameters can be found in table 1 of Ref.~\onlinecite{BernsteinPRE2020}. 
The results of this study show a novel effect on electrical transport is expected to influence these plasmas, which in turn influences other transport behaviors including magnetic field evolution and energy dissipation. 

%%%%%%%%
\section{Conductivity calculation~\label{sec:method}}

A simple model for electrical conduction can be derived using a first-order moment method that assumes both electrons and ions have Maxwellian distributions, but allows for a small drift of one species relative to the other. 
This corresponds to the first order of the Grad moment method of deriving transport properties, and is what is commonly cited in plasma formularies.~\cite{NRL1983}   
Here, we show how strong magnetization qualitatively changes electrical conduction using this common model. 

%%%%%%%%
\subsection{Force balance} 

The fluid force balance for species $s$ is
\begin{equation}
\label{eq:momentum} 
m_s n_s \frac{d \vc{V}_s}{dt} = n_s q_s (\vc{E} + \vc{V}_s \times \vc{B}) - \nabla p_s + \vc{R}^s , 
\end{equation}
where $\vc{R}^s$ is the friction force density on species $s$ and the pressure is that of an ideal gas $p_s = n_s T_s$. 
The magnetohydrodynamic (MHD) transport regime corresponds to the situation of small perturbations from equilibrium, so the plasma is quasineutral $n_i=n_e=n/2$ and electron and ion temperatures are nearly equal $T_e\approx T_i = T$. 
The steady-state force balance equation $\vc{J} \times \vc{B} = \nabla p$ follows from adding the electron and ion momentum equations from Eq.~(\ref{eq:momentum}) at steady-state, and identifying
\begin{equation}
\label{eq:J}
\vc{J} = \sum_s q_s n_s \vc{V}_s = e n_i (\vc{V}_i - \vc{V}_e)
\end{equation}
as the current density,
\begin{equation}
\vc{V} = \frac{n_i m_i \vc{V}_i + n_e m_e \vc{V}_e}{m_i n_i + m_e n_e} \approx \vc{V}_i 
\end{equation} 
as the fluid center of mass velocity, and making use of the approximation 
\begin{equation}
\vc{V}_e = \vc{V}_i - \frac{\vc{J}}{e n_i} \approx \vc{V} - \frac{\vc{J}}{en_i} 
\end{equation}
which is justified by the smallness of the electron-to-ion mass ratio. 
The total fluid pressure is $p=nT$, and the only non-zero component of the friction force density is the electron-ion drag $\vc{R}^{ie} (\vc{J}, \vc{B}) =- \vc{R}^{ei} (\vc{J}, \vc{B})$. 
A generalized Ohm's law can be written from the difference of the electron and ion momentum  balance equations at steady-state 
\begin{equation}
\label{eq:gen_ohm}
 \vc{E}^\prime = \frac{1}{en} \vc{J} \times \vc{B} - \frac{2 \vc{R}^{ie}}{en} ,
\end{equation} 
where $\vc{E}^\prime = \vc{E} + \vc{V} \times \vc{B}$. 
The remainder of this paper is concerned with the solution of Eq.~(\ref{eq:gen_ohm}) as the friction force density $\vc{R}^{ie}$ changes due to strong magnetization.  

%%%%%%%
\subsection{Electron-ion friction force} 

Typically, the friction force density is computed from the momentum moment of a collision operator $\vc{R}^{ie} = \int d^3v\, m_i \vc{v} C(f_i,f_e)$, where $C(f_i,f_e)$ describes interactions between the ion ($f_i$) and electron ($f_e$) distribution functions. 
Here, we present a method to compute $\vc{R}^{ie}$ from the sum of the average forces on individual particles. 
This provides an equivalent description for the model electron-ion friction problem that is a direct extension of the recent work on single particle motion in strongly magnetized plasmas.\cite{LafleurPPCF2019,LafleurPPCF2020,BernsteinPRE2020,JosePOP2020} 

The average trajectory of a test particle ($j$) in the presence of external electric and magnetic fields is described by the equation of motion
\begin{equation}
m_j \frac{d \vc{v}_j}{dt} = q_j (\vc{E} + \vc{v}_j \times \vc{B}) - \vc{F}_{j}
\end{equation}
where $\vc{F}_j$ is the average friction force on the test charge associated with its interaction with the rest of the plasma. 
This can be computed using linear response theory, which describes the electrostatic wake perturbations excited by the test charge from the linear Vlasov equation. 
The average force is then the product of the value of the test charge and the electrostatic field at the test charge location. 
The result can be expressed as~\cite{Ichimaru2004} 
\begin{equation}
\label{eq:Fj}
\vc{F}_j = - \frac{q_j^2}{\epsilon_o} \int \frac{d^3k}{(2\pi)^3} \frac{\vc{k}}{k^2} \textrm{Im} \biggl\lbrace \frac{1}{\hat{\varepsilon}(\vc{k}, \vc{k} \cdot \vc{v}_j)} \biggr\rbrace , 
\end{equation} 
where $\hat{\varepsilon}(\vc{k}, \omega)$ is the dielectric response function of the plasma. 
Here, we are interested in the interaction between electrons and ions. 
Considering a test ion, only the electron component of the dielectric response function is relevant to the drag due to electrons. 
Also, taking the electron distribution function to be a Maxwellian, the general linear plasma dielectric response can be expressed as~\cite{KrallPF1969} 
\begin{align}
\label{eq:full_epsilon}
\hat{\varepsilon}(\vc{k}, \omega) &= 1 + \frac{1}{k^2 \lambda_{De}^2} \biggl[ 1 + \frac{\omega}{|k_\parallel| v_{Te}} \exp \biggl( - \frac{k_\perp^2 v_T^2}{2 \omega_{ce}^2} \biggr) \\ \nonumber 
&\times \sum_{n=-\infty}^\infty I_n \biggl( \frac{k_\perp^2 v_{Te}^2}{2 \omega_{ce}^2} \biggr) Z \biggl( \frac{\omega - n \omega_{ce}}{|k_\parallel| v_{Te}} \biggr) \biggr] ,
\end{align} 
where $I_n$ is the $n$th-order modified Bessel function of the first kind, $Z$ is the plasma dispersion function, and $k_\parallel$ and $k_\perp$ are wavenumbers parallel and perpendicular to the magnetic field, respectively. 
In addition to the general formula from Eq.~(\ref{eq:full_epsilon}), we will also consider the limits of no magnetic field 
\begin{equation}
\label{eq:epsilon_o}
\hat{\varepsilon}_o(\vc{k}, \omega) = 1 + \frac{1}{k^2 \lambda_{De}^2} \biggl[ 1 + \frac{\omega}{kv_{Te}} Z \biggl( \frac{\omega}{kv_{Te}} \biggr) \biggr]
\end{equation}
and an arbitrarily strong magnetic field 
\begin{equation}
\label{eq:epsilon_inf}
\hat{\varepsilon}_\infty (\vc{k}, \omega) = 1 + \frac{1}{k^2 \lambda_{De}^2} \biggl[ 1 + \frac{\omega}{|k_\parallel | v_{Te}} Z \biggl( \frac{\omega}{|k_\parallel | v_{Te}} \biggr) \biggr] ,
\end{equation}
which follow directly from Eq.~(\ref{eq:full_epsilon}). 

As discussed in Ref.~\onlinecite{LafleurPPCF2019}, the general solution of Eqs.~(\ref{eq:Fj}) and (\ref{eq:full_epsilon}) for a test ion slowing on electrons is a vector that can be expressed as the sum of two components 
\begin{equation}
\vc{F}^{ie} = F^{ie}_v(v,\theta) \hat{\vc{v}} + F^{ie}_\times (v,\theta) \hat{\vc{v}} \times \hat{\vc{n}} ,
\end{equation}
where $\hat{\vc{n}} \equiv \hat{\vc{v}}\times \hat{\vc{B}}/\sin \theta$. 
The first component is antiparallel to the velocity vector of the test charge, and is commonly referred to as the stopping power~\cite{Nersisyan:2007}. 
In a strongly magnetized plasma, this is a function of both the speed of the particle $v$ and the angle between the velocity vector and the magnetic field $\theta = \arccos (\hat{\vc{v}} \cdot \hat{\vc{B}})$. 
The second component is perpendicular to the velocity vector in the plane of $\vc{v}$ and $\vc{B}$, and is referred to as the transverse force~\cite{LafleurPPCF2019}. 
This transverse component is comparable in magnitude to the stopping power in the strongly magnetized regime, but vanishes in the weakly magnetized (or zero field) regimes. 

The friction force density associated with the total ion distribution ($f_i$) is then obtained from the moment of the friction force on each ion, or equivalently, each infinitesimal slice of velocity phase-space
\begin{equation}
\label{eq:Rei_gen}
\vc{R}^{ie} = \int d^3v\, f_i(\vc{v}) \vc{F}^{ie}(\vc{v}) .
\end{equation} 
A simplification of this integral follows from making use of the fact that the velocity scale characterizing changes in $\vc{F}^{ie}$ is the electron thermal speed ($v_{Te}$). 
In contrast, the velocity scale characterizing $f_i$ is the much smaller ion thermal speed: $v_{Ti}/v_{Te} \sim \sqrt{m_e/m_i} \ll 1$. 
As a consequence, $f_i$ in Eq.~(\ref{eq:Rei_gen}) can be accurately approximated as a delta function
\begin{equation}
f_i(\vc{v}) = \frac{n_i}{\pi^{3/2} v_{Ti}^3} e^{-(\vc{v} - \vc{V}_i)^2/v_{Ti}^2} \approx n_i \delta (\vc{v} - \vc{V}_i)
\end{equation}
and the friction force density is approximately the product of the ion density and the average single particle friction evaluated at the ion speed: $\vc{R}^{ie} \approx n_i \vc{F}^{ie}(\vc{J})$. 
Since $\vc{V}_i$ is measured relative to $\vc{V}_e$ in this expression, this is simply a function of the current density $\vc{J}$: $\vc{V}_i \rightarrow \vc{V}_i - \vc{V}_e = \vc{J}/en_i$. 
The result is then a friction force density vector that is directly determined by the single-ion test particle problem 
\begin{equation}
\label{eq:Rie}
\vc{R}^{ie}(J,\theta) = n_i F_J^{ie}(J,\theta) \hat{\vc{J}} + n_i F_\times^{ie}(J,\theta) \hat{\vc{J}} \times \hat{\vc{n}}
\end{equation}
where now
\begin{equation}
\hat{\vc{n}} = \frac{\hat{\vc{J}} \times \hat{\vc{b}}}{\sin \theta} 
\end{equation}
and $\theta$ is the angle between $\vc{J}$ and $\vc{B}$ (which is the same as the angle between $\vc{v}$ and $\vc{B}$ in this approximation).

%%%%%%%%%
\subsection{Ohm's law} 

A generalized form of Ohm's law then follows from the combination of Eqs.~(\ref{eq:gen_ohm}) and (\ref{eq:Rie})
\begin{equation} 
\label{eq:ohm_nl}
\vc{E}^\prime = \frac{1}{en} \vc{J} \times \vc{B} + \eta_J \vc{J} + \eta_\times \vc{J} \times \hat{\vc{n}}, 
\end{equation} 
where 
\begin{equation}
\label{eq:eta_j}
\eta_J \equiv - \frac{F_J^{ie}(J,\theta)}{eJ}
\end{equation}
is the current-aligned resistivity coefficient and 
\begin{equation} 
\label{eq:eta_x}
\eta_\times \equiv - \frac{F_\times^{ie}(J,\theta)}{eJ}
\end{equation} 
is the transverse resistivity coefficient. 
Equation~(\ref{eq:ohm_nl}) is often rearranged considering two points of view. 
First, from the viewpoint of resistivity coefficients, the current and magnetic field vectors may be known, and Eq.~(\ref{eq:ohm_nl}) determines the electric field in the fluid reference frame ($\vc{E}^\prime$) that produces this current profile. 
For this problem, a convenient coordinate system is to take the magnetic field in the $\hat{z}$ direction ($\vc{B} = B \hat{z}$) and the current vector in the $x-z$ plane ($\vc{J} = J_x \hat{x} + J_z \hat{z}$). 
In this case, the components of Eq.~(\ref{eq:ohm_nl}) are
\begin{subequations}
\label{eq:resistivity}
\begin{eqnarray}
E^\prime_x &=& \eta_J J_x + \eta_\times J_z \\ 
E^\prime_y &=& - \frac{B}{en} J_x \\ 
E^\prime_z &=& \eta_J J_z - \eta_\times J_x  ,
\end{eqnarray}
\end{subequations}
providing
\begin{align}
\label{eq:ohm_wm}
\vc{E}^\prime &= \eta_\parallel \vc{J}_\parallel + \eta_\perp \vc{J}_\perp - \eta_\wedge \vc{J}_\wedge \\ \nonumber
&= \eta_\parallel J_z \hat{z} + \eta_\perp J_x \hat{x} - \eta_\wedge J_x \hat{y}
\end{align}
where
\begin{subequations}
\label{eq:eta_gen}
\begin{align}
\eta_\parallel &= \eta_J - \eta_\times \tan \theta \\ 
\eta_\perp &= \eta_J + \frac{\eta_\times}{\tan \theta} , \\
\eta_\wedge &= \frac{B}{en} .
\end{align}
\end{subequations}
Here, the parallel ($\parallel$), perpendicular ($\perp$), and polar ($\wedge$) directions are referenced with respect to the magnetic field. 
The $\eta_\wedge$ coefficient is the Hall resistivity. 

Second, from the viewpoint of conductivity, the electric and magnetic field vectors may be known, and Eq.~(\ref{eq:ohm_nl}) determines the current ($\vc{J}$) produced in response to the applied fields. 
For this problem, a convenient coordinate system is to take the magnetic field in the $\hat{z}$ direction ($\vc{B} = B \hat{z}$) and the electric field in the $\chi-z$ plane ($\vc{E}^\prime = E_\chi^\prime \hat{\chi} + E_z^\prime \hat{z}$). 
In this case, the components of Eq.~(\ref{eq:ohm_nl}) are
\begin{subequations}
\label{eq:conductivity}
\begin{eqnarray}
E^\prime_\chi &=& \frac{B}{en} J_\gamma + \eta_J J_\chi + \eta_\times \frac{J_\chi}{J_\perp} J_z \\ 
0 &=& - \frac{B}{en} J_\chi + \eta_J J_\gamma + \eta_\times \frac{J_\gamma}{J_\perp} J_z \\ 
E^\prime_z &=& \eta_J J_z - \eta_\times J_\perp .
\end{eqnarray}
\end{subequations}
where $J_\perp = \sqrt{J_\chi^2 + J_\gamma^2} = J \sin \theta$. 
%\begin{subequations}
%\label{eq:conductivity}
%\begin{eqnarray}
%E^\prime_\chi &=& \frac{B}{en} J_\gamma + \eta_\perp J_\chi  \\ 
%0 &=& - \frac{B}{en} J_\chi + \eta_\perp J_\gamma \\ 
%E^\prime_z &=& \eta_\parallel J_z ,
%\end{eqnarray}
%\end{subequations}
Identifying $\eta_\parallel$ and $\eta_\perp$ coefficients, and inverting the result provides 
\begin{align}
\label{eq:ohm_wm_2}
\vc{J} &= \sigma_\parallel \vc{E}^\prime_\parallel + \sigma_\perp \vc{E}^\prime_\perp + \sigma_\wedge \vc{E}^\prime_\wedge \\ \nonumber
&= \sigma_\parallel E^\prime_z \hat{z} + \sigma_\perp E^\prime_\chi \hat{\chi} + \sigma_\wedge E^\prime_\chi \hat{\gamma}
\end{align}
where
\begin{subequations}
\label{eq:sig_gen}
\begin{align}
\sigma_\parallel &= \frac{1}{\eta_\parallel}, \\ 
\sigma_\perp &= \frac{\eta_\perp}{\eta_\wedge^2 + \eta_\perp^2} , \\
\sigma_\wedge &= \frac{\eta_\wedge}{\eta_\wedge^2 + \eta_\perp^2} 
\end{align}
\end{subequations}
are the parallel, perpendicular, and Hall conductivity coefficients computed in terms of the resistivity coefficients. 
The next two sections explore the solutions of Eqs.~(\ref{eq:eta_gen}) and (\ref{eq:sig_gen}) in the weak magnetization and strong magnetization regimes.  
% $J_\perp = \sqrt{J_\chi^2 + J_\gamma^2} = J \sin \theta$

%%%%%%%%%
\section{Weakly magnetized} 

In the limit that $\beta_e \ll 1$, the general plasma dielectric response function from Eq.~(\ref{eq:full_epsilon}) is accurately approximated by the low field limit from Eq.~(\ref{eq:epsilon_o}). 
Appendix~\ref{ap:r_wm} shows that taking this limit, and also assuming that $v \ll v_{Te}$, Eq.~(\ref{eq:Fj}) reduces to 
\begin{equation}
\label{eq:Fj_wm}
\vc{F}_j^\textrm{w} = - \frac{e^4 2 \sqrt{2\pi} \sqrt{m_e} n \ln \Lambda}{(4\pi \epsilon_o)^2 3 (k_BT)^{3/2}} \vc{v}_j .
\end{equation}
Here, the superscript ``w'' denotes the weakly magnetized limit, and $j$ refers to a single particle. 
Applying Eq.~(\ref{eq:Fj_wm}) to Eqs.~(\ref{eq:eta_j}) and~(\ref{eq:eta_x}) and making use of the definition of current density from Eq.~(\ref{eq:J}) provides 
\begin{subequations}
\label{eq:eta_w1}
\begin{align}
\eta_J^\textrm{w} &= \eta_o, \\
\eta_\times^\textrm{w} &= 0.
\end{align}
\end{subequations}
where 
\begin{equation}
\label{eq:eta_o}
\eta_o = \frac{m_e}{ne^2 \tau_e}
\end{equation}
is the commonly understood reference plasma resistivity~\cite{NRL1983}. 
As expected, Eq.~(\ref{eq:eta_w1}) returns the result that is usually derived from the collision operator of the plasma kinetic equation, demonstrating that the independent method applied here is equivalent to the more common method based on a collision operator~\cite{NRL1983}.  
%This is a constant resistivity coefficient that is independent of $\vc{B}$ and $\vc{J}$, and 
The transverse resistivity coefficient is zero $\eta_\times^\textrm{w} = 0$.

%%%%%%
\subsection{Electrical resistivity} 

In the weakly magnetized limit, the resistivity coefficients from Eq.~(\ref{eq:eta_w1}) reduce to a set of linear equations with constant coefficients in the coordinate system of Eq.~(\ref{eq:eta_gen})
\begin{subequations}
\label{eq:eta_w}
\begin{eqnarray}
\eta_\parallel^\textrm{w} &=& \eta_o \\
\eta_\perp^\textrm{w} &=& \eta_o \\
\eta_\wedge^\textrm{w} &=& \frac{B}{en} =  \eta_o \omega_{ce} \tau_e .
\end{eqnarray}
\end{subequations}
Here, 
\begin{equation}
\label{eq:Hall}
\omega_{ce} \tau_e = \sqrt{\frac{3\pi}{8}} \frac{\beta_e}{\Gamma_e^{3/2} \ln \Lambda} 
\end{equation}
is the Hall parameter, which quantifies if a plasma is magnetized in the sense that the magnetic field influences collisional transport.

The traditional first-order moment method, or first-order Chapman-Enskog solution, of the plasma kinetic equation corresponds to the solution of Eq.~(\ref{eq:ohm_nl}) with $\eta_J = \eta_o$ and $\eta_\times = 0$. 
Equation~(\ref{eq:eta_w1}) returns the precisely same result, and leads to the equal parallel and perpendicular resistivity coefficients in Eq.~(\ref{eq:eta_w}). 
Inequality of $\eta_\parallel$ and $\eta_\perp$ in the Braginskii transport equations arises at higher order in the expansion of the distribution functions away, as the electron distribution function exhibits an anisotropy with respect to the magnetic field~\cite{Ferziger1972}.  
For comparison, the second-order solution is provided in Appendix~\ref{app:ce}. 
This anisotropy between $\eta_\parallel$ and $\eta_\perp$ at second order also implies a non-zero transverse resistivity $\eta_\times \neq 0$. 
This is discussed further in Sec.~\ref{sec:sm} and Appendix~\ref{app:ce}. 
Here, we focus only on the first order solution to provide a commensurate comparison with the solution for a strongly magnetized plasma computed at the same order.

%%%%%%%
\subsection{Electrical conductivity} 

%Since the relationship between $\vc{E}^\prime$ and $\vc{J}$ is linear in the weakly magnetized regime, the coefficients in 
%Eq.~(\ref{eq:ohm_wm}) can be easily inverted to provide the current density $\vc{J}$ associated with a given electric field $\vc{E}^\prime$. 
Utilizing the resistivity coefficients from Eq.~(\ref{eq:eta_w}), the electrical conductivity coefficients from Eq.~(\ref{eq:sig_gen}) are 
\begin{subequations}
\label{eq:sigma_w}
\begin{align}
\sigma_\parallel^{\textrm{w}} &= \sigma_o \\
\sigma_\perp^\textrm{w} &=  \frac{\sigma_o}{1 + (\omega_{ce} \tau_e)^2} \\
\sigma_\wedge^\textrm{w} &= \frac{\sigma_o}{\omega_{ce} \tau_e} \frac{(\omega_{ce} \tau_e)^2}{1 + (\omega_{ce} \tau_e)^2} ,
\end{align}
\end{subequations} 
where $\sigma_o = 1/\eta_o$ is a reference conductivity coefficient. 
Again, this corresponds exactly to the first-order Chapman-Enskog solution. 
For comparison, the second order solution is provided in the Appendix~\ref{app:ce}. 
Equations~(\ref{eq:eta_w}) and (\ref{eq:sigma_w}) are reproduced here in order to show that the method of Sec.~\ref{sec:method} reproduces the well known results. 
Next, we apply the same method to the limit of strong magnetization to show how these solutions change.

%%%%%%%%%%
\section{Strongly Magnetized\label{sec:sm}} 
%\subsection{Electron-ion friction force} 

In the limit that $\beta_e \gtrsim \Gamma_e^{-3/2}$, the general plasma dielectric response function from Eq.~(\ref{eq:full_epsilon}) is accurately approximated by the infinite field limit from Eq.~(\ref{eq:epsilon_inf})~\cite{LafleurPPCF2019}. 
Taking this limit, and also assuming that $v \ll v_{Te}$ and $\Gamma_e \ll 1$, Appendix~\ref{app:beta_inf} shows that Eq.~(\ref{eq:Fj}) reduces to 
\begin{equation}
\label{eq:Fj_s}
\vc{F}_j^\textrm{s} = F_v^{\textrm{s}} \hat{\vc{v}}_j + F_\times^{\textrm{s}} \hat{\vc{v}}_j \times \hat{\vc{n}}
\end{equation}
where 
%\begin{subequations}
%\begin{align}
\begin{equation}
\label{eq:fv_s}
F_v^\textrm{s} = - \frac{e^4 \sqrt{2\pi} \sqrt{m_e} n v_j \ln \Lambda}{(4\pi \epsilon_o)^2 (k_BT)^{3/2}} \biggl[ \cos^2 \theta - \sin^2 \theta \ln \biggl( \frac{v_{j\perp} }{v_{Te}} \biggr) \biggr] 
\end{equation}
is the stopping power component, and
\begin{equation}
\label{eq:fx_s}
F_\times^\textrm{s} =  \frac{e^4 \sqrt{2\pi} \sqrt{m_e} n v_j \ln \Lambda}{(4\pi \epsilon_o)^2(k_BT)^{3/2}} \sin \theta \cos \theta \biggl[ 1 + \ln \biggl( \frac{v_{j\perp}}{v_{Te}} \biggr) \biggr] 
\end{equation}
is the transverse friction force. 
Here, $v_{j,\perp} = v_j | \sin \theta|$. 
%\end{align}
%\end{subequations} 
Equations~(\ref{eq:Fj_s})--(\ref{eq:fx_s}) can be interpreted via a generalized Coulomb collision frequency. 
Because the magnetic field does not influence the collision process in the weakly magnetized regime, there is only a single component of the friction force ($F_v$) and a single collision frequency provided by Eq.~(\ref{eq:tau_e}). 
In contrast, the influence of the magnetic field on collisions in a strongly magnetized regime causes an anisotropy in the collision frequency. 
This results in the introduction of the transverse force $F_\times$ associated with a collision frequency that is distinct from that which describes stopping power. 
It also introduces an angular ($\theta$) dependence into the collision frequencies associated with each of these vector components. 

Applying Eq.~(\ref{eq:Fj_s}) to Eqs.~(\ref{eq:eta_j}) and~(\ref{eq:eta_x}), and making use of the definition of current density from Eq.~(\ref{eq:J}), provides the resistivity coefficients 
\begin{subequations}
\label{eq:eta_strong}
\begin{align}
\label{eq:eta_j_2}
\eta_J^\textrm{s} &= \eta_o \frac{3}{2} [ \cos^2 \theta - \sin^2 \theta \ln (J_\perp/J_o)] , \\
\label{eq:eta_times}
\eta_\times^\textrm{s} &= - \eta_o \frac{3}{2} \sin \theta \cos \theta [1 + \ln (J_\perp/J_o)] 
\end{align}
\end{subequations}
where $J_o \equiv en_i v_{Te}$ and $J_\perp = J |\sin \theta|$. 
Note that since $J_\perp/J_o \ll 1$, the second term of Eq.~(\ref{eq:eta_j_2}) always has a positive sign. 
Here, the superscript $s$ denotes the strongly magnetized limit. 
The generalized Ohm's law described by Eqs.~(\ref{eq:ohm_nl}), (\ref{eq:eta_j_2}) and (\ref{eq:eta_times}) differs substantially from that obtained in the weakly magnetized limit [Eq.~(\ref{eq:eta_w1})]. 

A fundamental difference is that it becomes nonlinear due to the logarithmic dependence on the current density. 
This nonlinearily results from the $M\sin \theta \ln (M \sin \theta)$ dependence of the friction force in the $x$-direction in the limit $M \ll 1$: see Eq.~(\ref{eq:Fx}). 
Although this implies that the resistivity coefficients diverge logarithmically as $J_\perp \rightarrow 0$, it is not an unphysical effect since the observable electric field results from the product of the resistivity coefficients with a linear factor of the perpendicular current, and  $\lim_{J_\perp \rightarrow 0} J_\perp \ln J_\perp =0$. 
The next two subsections will also show that this logarithmic nonlinearity is weak, and the dependence on $J_\perp$ can be accurately approximated by a factor of the Hall parameter $\omega_{ce} \tau_e$: $\ln(J_\perp/J_o) \approx - \ln (\omega_{ce} \tau_e)$. 
This leads to the result 
\begin{subequations}
\label{eq:eta_strong_3}
\begin{align}
\label{eq:eta_j_3}
\eta_J^\textrm{s} &\approx \eta_o \frac{3}{2} [ \cos^2 \theta + \sin^2 \theta \ln (\omega_{ce} \tau_e)] , \\
\label{eq:eta_times_3}
\eta_\times^\textrm{s} &\approx - \eta_o \frac{3}{2} \sin \theta \cos \theta [1 - \ln (\omega_{ce} \tau_e)] .
\end{align}
\end{subequations}

Beyond the nonlinearity, strong magnetization causes the resistivity coefficients to depend on the orientation of the current with respect to the magnetic field. 
It also causes there to be a non-zero transverse resistivity coefficient. 
Here, both of these effects result from an asymmetry in the Coulomb collision rate between the directions parallel and perpendicular to the magnetic field. 
Neither effect is observed when the weakly magnetized collision model is applied to the same first-order moment method, as demonstrated by Eq.~(\ref{eq:eta_w1}). 
It should be noted that the resistivity coefficients computed at second-order in the Chapman-Enskog do exhibit an anisotropy; see Appendix~\ref{app:ce}. 
However, it is for a different reason. 
The second order Chapman-Enskog solution accounts for the distortion of the electron distribution function away from equilibrium, and the restoring effect from electron-electron collisions. 
In this situation, asymmetry arises from the Lorentz term of the convective derivative in the kinetic equation, which leads to an effect that the electron velocity distribution function is more easily distorted along the magnetic field than across it in response to the electric field. 
In this case, the collision physics itself is not influenced by the magnetic field. 
In contrast, the asymmetry that arises in Eqs.~(\ref{eq:eta_j_2}) and (\ref{eq:eta_times}) is due to an asymmetry in the rate of Coulomb collisions that occurs at the microscopic (sub-Debye length) scale.

%%%%%%
\subsection{Electrical resistivity} 

The resistivity coefficients from Eq.~(\ref{eq:eta_strong}) can be cast in the coordinate system defined by Eq.~(\ref{eq:eta_gen}). 
This provides the coefficients
\begin{subequations}
\label{eq:eta_s_mid}
\begin{align}
\eta_\parallel^\textrm{s} &= \frac{3}{2} \eta_o \\
\label{eq:eta_perp_s} 
\eta_\perp^\textrm{s} &= - \frac{3}{2} \eta_o \ln \biggl( \frac{J_\perp}{J_o}\biggr) \\ 
\eta_\wedge^\textrm{s} &=  \frac{B}{en} =  \eta_o \omega_{ce} \tau_e .
\end{align}
\end{subequations}
Comparing with the results in the weakly magnetized limit from Eq.~(\ref{eq:eta_w}), strong magnetization changes $\eta_\parallel$ only by the constant numerical factor $3/2$. 
This enhanced resistivity is associated with the enhanced friction force observed at low speeds due to strong magnetization for particles with velocities aligned along the magnetic field. 
For example, this can be observed in figure 3a of Ref.~\onlinecite{LafleurPPCF2019}. 
Because it is not influenced by collisions, the Hall resistivity is unchanged. 
The most significant effect is observed in the perpendicular direction. 
In addition to the $3/2$ factor, strong magnetization is predicted to enhance the perpendicular resistivity by a potentially large logarithmic factor. 

Although the logarithmic factor in Eq.~(\ref{eq:eta_perp_s}) depends on the perpendicular current, this nonlinearity is weak and the current dependence can be replaced by the Hall parameter. 
In particular, inverting the resistive form of Ohm's law to obtain conductivity coefficients (see Sec.~\ref{sec:sm_sigma}) will show that in the limit that $\omega_{ce} \tau_e \gg \ln (\omega_{ce} \tau_e)$, $J_\perp \approx E_\chi^\prime/[\eta_o(\omega_{ce}\tau_e)]$, so that $\ln (J_\perp/J_o) \approx \ln[\bar{E}_\chi^\prime/(\eta_o J_o)] - \ln (\omega_{ce} \tau_e) \approx - \ln (\omega_{ce} \tau_e)$. 
Since we are working in the limit of strong magnetization here ($\beta_e \gtrsim \Gamma_e^{-3/2}$), and Eq.~(\ref{eq:Hall}) provides $\omega_{ce} \tau_e \propto \Gamma_e^{-3/2} \beta_e \gtrsim \Gamma_e^{-3}$. 
In weakly coupled plasmas, $\Gamma_e \ll 1$ is a very small number, so it is expected that $\omega_{ce} \tau_e$ must be very large to reach the strongly magnetized limit. 
The assumption of a weak nonlinearity is well justified in this limit. 
With this, Eq.~(\ref{eq:eta_s_mid}) simplifies to the final result for the predicted resistivity coefficients in the strongly magnetized regime 
\begin{subequations} 
\label{eq:eta_sum_2}
\begin{align}
\eta^\textrm{s}_\parallel &\approx \frac{3}{2} \eta_o , \\ 
\eta^\textrm{s}_\perp &\approx \frac{3}{2}\eta_o \ln (\omega_{ce} \tau_e) , \\
\eta^\textrm{s}_\wedge &\approx \eta_o \omega_{ce} \tau_e . 
\end{align}
\end{subequations} 
which are also summarized in Eq.~(\ref{eq:eta_sum}). 
By comparing Eq.~(\ref{eq:eta_sum_2}) with Eq.~(\ref{eq:eta_w}), the most significant change arising from strong magnetization is a significant enhancement of the perpendicular resistivity, by a factor of $\frac{3}{2} \ln (\omega_{ce} \tau_e) \gg 1$.

%%%%%%%
\subsection{Electrical conductivity\label{sec:sm_sigma}}  

The electrical conductivity coefficients are obtained using Eqs.~(\ref{eq:sig_gen}) and (\ref{eq:eta_s_mid}), which provides 
\begin{subequations}
\label{eq:sigma_s}
\begin{align}
\sigma_\parallel^{\textrm{s}} &= \frac{2}{3} \sigma_o \\
\sigma_\perp^\textrm{s} &= \sigma_o  \frac{[- \frac{3}{2} \ln (J_\perp/J_o)]}{[- \frac{3}{2} \ln (J_\perp/J_o)]^2 + (\omega_{ce} \tau_e)^2} \\
\sigma_\wedge^\textrm{s} &=  \frac{\sigma_o}{\omega_{ce} \tau_e} \frac{(\omega_{ce} \tau_e)^2}{[- \frac{3}{2} \ln (J_\perp/J_o)]^2 + (\omega_{ce} \tau_e)^2} .
\end{align}
\end{subequations} 
Again, the conductivity coefficients are predicted to depend on a logarithmic nonlinearity in the perpendicular current. 
However, we can now justify the previous argument that this nonlinearity is weak in the strongly magnetized regime. 

Specifically, note that $J_\perp = \sqrt{J_\chi^2 + J_\gamma^2}$ gives
\begin{equation}
J_\perp = \frac{E_\chi^\prime/\eta_o}{\sqrt{[- \frac{3}{2} \ln (J_\perp/J_o)]^2 + (\omega_{ce} \tau_e)^2}} \approx \frac{E_\chi^\prime}{\eta_o (\omega_{ce} \tau_e)} 
\end{equation}
since $\omega_{ce} \tau_e \gg [- \frac{3}{2} \ln (\bar{J}_\perp)]$. 
Thus,  $\ln (J_\perp/J_o) \approx \ln[\bar{E}_\chi^\prime/(\eta_o J_o)] - \ln (\omega_{ce} \tau_e) \approx - \ln (\omega_{ce} \tau_e)$, which is the argument utilized above to reduce Eq.~(\ref{eq:eta_s_mid}) to Eq.~(\ref{eq:eta_sum_2}).  
With this approximation, we arrive at the final result for the predicted conductivity coefficients in the strongly magnetized regime 
\begin{subequations}
\label{eq:sig_sum_2}
\begin{align}
\sigma_\parallel^\textrm{s} & \approx \frac{2}{3} \sigma_o ,\\
\sigma_\perp^\textrm{s} &\approx \frac{3}{2} \sigma_o \frac{\ln (\omega_{ce} \tau_e)}{(\omega_{ce} \tau_e)^2} ,\\
\sigma_\wedge^\textrm{s} &\approx  \frac{\sigma_o}{\omega_{ce} \tau_e} .
\end{align}
\end{subequations} 
Similar to the resistivity coefficients, strong magnetization is found to lead to a slight change to the parallel conductivity (by a $2/3$ factor in this case), but the most significant change is an enhancement of the perpendicular conductivity by a factor of $\frac{3}{2} \ln (\omega_{ce} \tau_e)$. 

%%%%%%%%%%%%
\section{Connection formula} 

The previous two sections developed analytic solutions in the weakly and strongly magnetized limits. 
In principle, the general case may be treated by solving Eq.~(\ref{eq:Fj}) with the dielectric response from Eq.~(\ref{eq:full_epsilon}) numerically. 
Although the friction force on a test ion can, and has~\cite{LafleurPPCF2019}, been made in this way, obtaining a generalized Ohm's law of the form obtained in the previous two sections is not likely to be possible. 
This is because the friction force depends nonlinearly on the speed, even in the low speed limit, in a way that is difficult to determine. 
Instead, we assume that the weakly and strongly magnetized limits are joined continuously, and match the connection between these regimes using a Pad\'{e} approximation.

\begin{figure*}
\includegraphics[width=16cm]{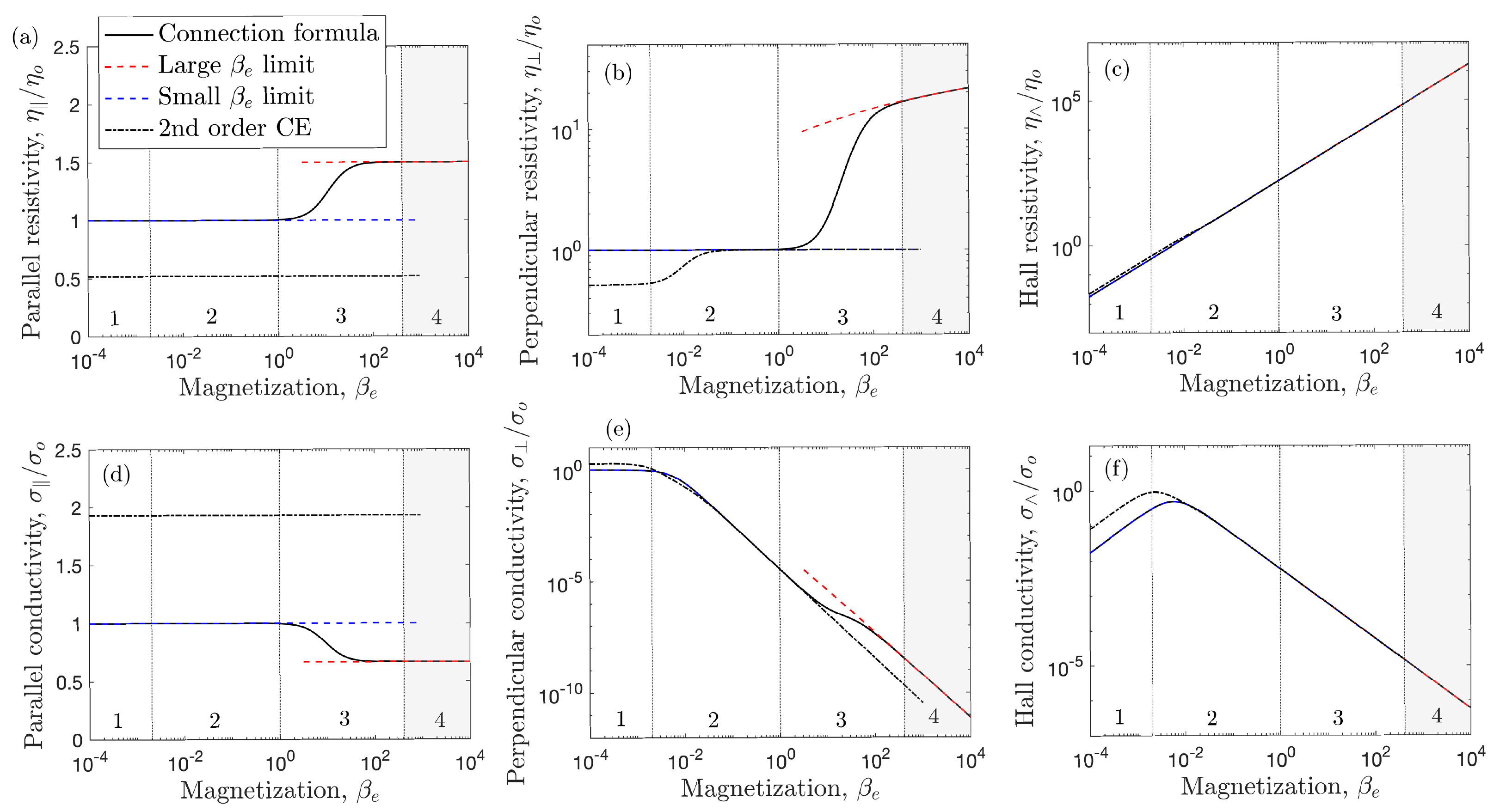}
\caption{Parallel (a), perpendicular (b), and Hall (c) resistivity coefficients, as well as parallel (d), perpendicular (e), and Hall (f) conductivity coefficients, plotted as a function of the magnetization parameter $\beta_e$. 
Connection formulas from Eq.~(\ref{eq:eta_connect}) span the magnetization regimes (black lines), whereas those from Eqs.~(\ref{eq:eta_sum_2}) and (\ref{eq:sig_sum_2}) give the large $\beta_e$ limit (red dashed lines) and those from Eqs.~(\ref{eq:eta_w}) and (\ref{eq:sigma_w}) give the small $\beta_e$ limit (blue dashed lines). 
The second-order Chapman Enskog solutions for the weakly magnetized regimes from Eqs.~(\ref{eq:eta_2_CE}) and (\ref{eq:sigma_CE}) are also shown (dash-dotted line). 
Vertical dashed lines denote the separation of the four transport regimes described in the text: (1) unmagnetized, (2) magnetized, (3) strongly magnetized, and (4) extremely magnetized. 
Region 4 is shaded because the theoretical approach used is expected to begin to break down in that region. 
The Coulomb coupling strength was chosen to be $\Gamma_e = 10^{-2}$. 
}
\label{fg:eta_fig}
\end{figure*}

To do so, we first require the second order solution in the $\beta_e$ expansion for the resistivity coefficients. 
Applying the results from Eqs.~(\ref{eq:Fz_wb}) and (\ref{eq:Fx}), we have to second order in the $\beta_e \ll 1$ expansion
\begin{subequations}
\begin{align}
\eta_\parallel^{\textrm{w}2} &= \eta_o \bigl( 1 + \frac{\beta_e^2}{40\ln \Lambda} \bigr) , \\
\eta_\perp^{\textrm{w}2} &= \eta_o \bigl(1 + \frac{\beta_e^2}{20 \ln \Lambda} \bigr) , \\
\eta_\wedge^{\textrm{w}2} &=  \eta_o (\omega_{ce} \tau_e) .
\end{align} 
\end{subequations}
Using the result that the second order correction is $\propto \beta_e^2$, and the results from Eq.~(\ref{eq:eta_sum_2}) for the large $\beta_e$ asymptotic limit, the Pad\'{e} approximate is found by assuming a solution of the form $\eta/\eta_o \propto (1+a \beta_e^2)/(1+b \beta_e^2)$ and solving for the coefficients $a$ and $b$ by fitting both the series expansion for $\beta_e \ll 1$ and the asymptotic expansion for $\beta_e \rightarrow \infty$ to these known limits. 
The process is repeated for $\eta_\parallel$ and $\eta_\perp$, resulting in 
\begin{subequations}
\begin{align}
\eta_\parallel &= \eta_o \frac{40 \ln \Lambda + 3 \beta_e^2}{40 \ln \Lambda + 2 \beta_e^2} , \\
\eta_\perp &= \eta_o \frac{40 \ln \Lambda [\frac{3}{2} \ln (\omega_{ce} \tau_e) -1] + 3 \ln (\omega_{ce} \tau_e) \beta_e^2}{40 \ln \Lambda [\frac{3}{2} \ln (\omega_{ce} \tau_e) - 1] + 2 \beta_e^2} \\ 
\eta_\wedge &= \eta_o (\omega_{ce} \tau_e) .
\end{align}
\end{subequations}
This can be written in terms of the Hall parameter ($\omega_{ce} \tau_e$) instead of $\beta_e$ by making use of Eq.~(\ref{eq:Hall}), providing the form
\begin{subequations}
\label{eq:eta_connect}
\begin{align}
\eta_\parallel &= \eta_o \frac{15 \pi + 3 \Gamma_e^3 \ln \Lambda (\omega_{ce} \tau_e)^2}{15 \pi + 2 \Gamma_e^3 \ln \Lambda (\omega_{ce} \tau_e)^2} , \\
\eta_\perp &= \eta_o \frac{15\pi [\frac{3}{2} \ln (\omega_{ce} \tau_e) -1] + 3 \Gamma_e^3 \ln \Lambda \ln (\omega_{ce} \tau_e) (\omega_{ce} \tau_e)^2}{15 \pi [\frac{3}{2} \ln (\omega_{ce} \tau_e) - 1] + 2 \Gamma_e^3 \ln \Lambda (\omega_{ce} \tau_e)^2} ,\\
\eta_\wedge &= \eta_o (\omega_{ce} \tau_e) .
\end{align}
\end{subequations}
This form retains the known limit to second order for $\beta_e \ll 1$, as well as the asymptotic limit for $\beta_e \rightarrow \infty$, and smoothly connects them through the intermediate $\beta_e$ regime. 
Connection formula for the conductivity coefficients are obtained directly by inserting the results of Eq.~(\ref{eq:eta_connect}) into Eq.~(\ref{eq:sig_gen}). 

Plots of the resistivity and conductivity coefficients obatined from Eq.~(\ref{eq:eta_connect}) are shown in Fig.~\ref{fg:eta_fig}, along with the low magnetic field limit from Eqs.~(\ref{eq:eta_w}) and (\ref{eq:sigma_w}), the large magnetic field limit from Eqs.~(\ref{eq:eta_sum_2}) and (\ref{eq:sig_sum_2}), and the second-order solution of the Chapman-Enskog solution of the plasma kinetic equation for the weakly magnetized limit from Eqs.~(\ref{eq:eta_2_CE}) and (\ref{eq:sigma_CE}). 
In this figure, the Coulomb coupling strength was chosen to be $\Gamma_e = 10^{-2}$. 

The four regimes are delineated on the figure correspond to the boundaries at which dimensionless length scales transition in such a way that different physical processes are expected to control transport. 
These were previously identified in Ref.~\onlinecite{BaalrudPRE2017}, where they were tested using molecular dynamics simulations: (1) Unmagnetized ($r_c > \lambda_{\textrm{coll}}$): In this region the gyroradius is less than the Coulomb collision mean free path ($\lambda_{\textrm{coll}} = v_{Te} \tau_e$). 
Here, magnetization is not expected to influence transport. 
(2) Magnetized ($\lambda_{D} < r_c < \lambda_{\textrm{coll}}$): Here, magnetization influences macroscopic transport due to the influence of the Lorentz force on the distribution functions, but it does not influence Coulomb collisions at the microscopic (sub-Debye length) scale. 
This is the regime of traditional plasma kinetic, and Braginskii transport theory. 
(3) Strongly magnetized ($r_L < r_c < \lambda_{D}$): Here, magnetization influences both the evolution of the distribution function on macroscopic scales, and Coulomb collisions at microscopic scales, but the gyroradius remains larger than the distance of closest approach in a binary collision. 
This is the regime in which the linear response theory described above is expected to predict significant changes due to gyromotion at the collision scale. 
(4) Extremely magnetized ($r_c < r_L$): Here, magnetization is so strong that the gyroradius is the smallest length scale relevant to collisions. 
The linear reponse theory is not expected to apply in this regime because large-angle close interactions come to dominate transport~\cite{JosePOP2020,VidalPOP2021}. 
Here, collisional kinetic theories based on a generalization of the Boltzmann-type approach, such as that recently developed in Ref.~\onlinecite{JosePOP2020}, are required. 

Figure~\ref{fg:eta_fig} shows the expected behavior from standard theory in regions 1 and 2: The parallel and perpendicular resistivity coefficients are independent of the field strength, and the perpendicular conductivity coefficient falls off as $\beta_e^{-2}$. 
The new changes due to strong magnetization arise in region 3: Parallel resistivity and conductivity change with magnetic field strength, the perpendicular resistivity is significantly enhanced, and the perpendicular conductivity is enhanced. 
It should be noted that the connection formula in region 3 are simply based on matching the asymptotic limits that could be calculated analytically. 
It is expected to be accurate through regions 1 and 2, and in the transition to region 3, as well as the transition to region 4. 
The functional form of the curve is not necessarily trustworthy in the middle of region 3. 
It also does not apply for $\beta_e$ well into region 4 (only in the transition to this region).

%%%%%%%%%%%%
\section{Conclusions} 

This work shows that magnetization at the scale of collisions translates to qualitative changes in the electrical resistivity and conductivity coefficients. 
It causes the parallel coefficient to depend on the magnetic field strength, and the perpendicular coefficients to increase substantially. 
These changes come about when the electron gyrofrequency exceeds the electron plasma frequency, which corresponds to the conditions at which the electron gyroradius is less than the Debye length. 
The approach presented in this work was limited to describing the contribution to electrical conduction from electron-ion interactions. 
A more complete account that includes the electron-electron interactions will require a different solution method, such as a Chapman-Enskog solution of the generalized Boltzmann equation for strongly magnetized plasmas.~\cite{JosePOP2020} 

Although this work focused on electrical conduction, the main result that strong magnetization influences transport rates in qualitatively new ways is also expected to apply to other transport processes, such as thermal conductivity~\cite{HollmannPRL1999,OttPRE2015}, viscosity~\cite{KrieselPRL2001,ScheinerPRE2020}, diffusion~\cite{OttPRL2011,BaalrudPRE2017}, and temperature relaxation~\cite{BeckPOP1996,OttPRE2015}. 
Plasmas that reach strong magnetization regimes, such as non-neutral and ultracold plasmas, astrophysical plasmas, and even to some extent fusion plasmas, are expected to be influenced by these changes. 
Further work will be required to develop a comprehensive magnetohydrodynamic description for strongly magnetized plasmas. 

\section{Data Availability} 

The data that support the findings of this study are available from the corresponding author upon reasonable request.

%%%%%%%
\appendix

%%%%%%%%%%
\section{Comparison of first and second order Chapman-Enskog solutions\label{app:ce}} 

For comparison with the test particle method, we briefly recall the results of the second order Chapman-Enskog method. 
These can be found in classic works such as Braginskii~\cite{BraginskiiRPP1965}, or textbooks~\cite{Chapman:1991,Ferziger1972}. 
%In this model the source of differences in the parallel and perpendicular resistivity and conductivity coefficients is the convective (Lorentz) term of the kinetic equation, rather than the collision operator. 
To second order, the resistivity coefficients are
\begin{widetext}
\begin{subequations}
\label{eq:eta_CE}
\begin{eqnarray}
\eta_\parallel^\textrm{CE} &=& 0.518 \eta_o \\ 
\eta_\perp^\textrm{CE} &=& \eta_o \frac{(\omega_{ce} \tau_e)^6 + 8.084 (\omega_{ce} \tau_e)^4 + 12.25 (\omega_{ce}\tau_e)^2 + 1.681}{(\omega_{ce} \tau_e)^6 + 9.764 (\omega_{ce}\tau_e)^4 + 22.81 (\omega_{ce}\tau_e)^2 + 3.247} \\
\eta_\wedge^\textrm{CE} &=& -\eta_o (\omega_{ce} \tau_e) \frac{(\omega_{ce}\tau_e)^6 + 10.66 (\omega_{ce}\tau_e)^4 + 28.46 (\omega_{ce}\tau_e)^2 + 4.088}{(\omega_{ce}\tau_e)^6 + 9.764(\omega_{ce}\tau_e)^4 + 22.81(\omega_{ce}\tau_e)^2 + 3.247 } .
\end{eqnarray}
\end{subequations}
\end{widetext}
The distinction between the simplified model results from Eqs.~(\ref{eq:eta_w}) and the more comprehensive second-order Chapman-Enskog solution from Eq.~(\ref{eq:eta_CE}) is the contribution from the distortion of the electron velocity distribution function away from a Maxwellian, and the associated electron-electron interactions. 

\begin{figure}
\includegraphics[width=8cm]{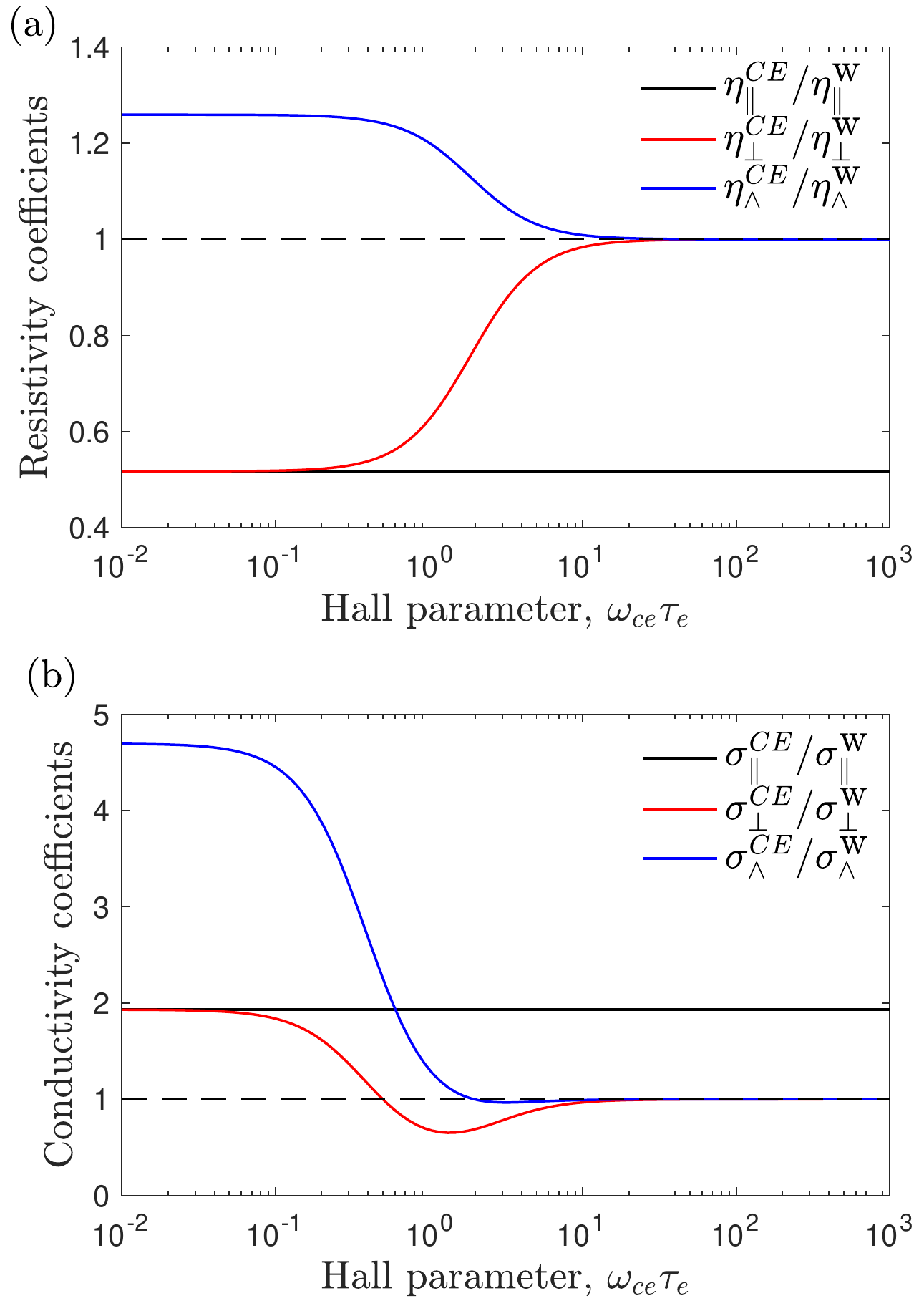}
\caption{(a) Ratio of the resistivity coefficients obtained from the second-order Chapman-Enskog method from Eq.~(\ref{eq:eta_CE}) and the first-order moment method from Eq.~(\ref{eq:eta_w}) as a function of the Hall parameter $\omega_{ce} \tau_e$. 
(b) Ratio of the conductivity coefficients obtained from the second-order Chapman-Enskog method from Eq.~(\ref{eq:sigma_CE}) and the first-order moment method from Eq.~(\ref{eq:sigma_w}) as a function of the Hall parameter $\omega_{ce} \tau_e$. 
%The MD data clearly supports the identified boundaries as locations where transport properties qualitatively change between regimes.  
}
\label{fg:CE}
\end{figure}

Figure~\ref{fg:CE} shows a comparison of the results of the two methods. 
This demonstrates that all coefficients are within a factor of two from each method. 
The largest difference is the Spitzer correction (factor of 0.518) in the parallel coefficient, which is independent of magnetic field strength, and is directly associated with the electron contribution. 
Perpendicular resistivity is unaffected by the Spitzer correction in the magnetized regime ($\omega_{ce} \tau_e \gg 1$), but asymptotes to the parallel coefficient in the unmagnetized regime. 
Similarly, the Hall resistivity is identical for each method in the magnetized regime ($\omega_{ce} \tau_e \gg 1$), but differs by approximately 25\% in the unmagnetized regime. 
In this work, we are primarily interested in understanding the transition from magnetized to strongly magnetized plasma that happens at $\beta_e > 1$. 
The magnetization parameter is related to the Hall parameter by Eq.~(\ref{eq:Hall}). 
%\begin{equation}
%\beta_e = \sqrt{\frac{3\pi}{8}} \frac{\omega_{ce} \tau_e}{\Gamma_e^{3/2} \ln \Lambda} .
%\end{equation} 
Since we only consider weakly coupled plasmas ($\Gamma_e \ll 1$), the magnetization parameter is much larger than the Hall parameter $\beta_e \gg \omega_{ce} \tau_e$. 
For this reason, we will focus on the $\omega_{ce} \tau_e \gg 1$ regime, where Fig.~\ref{fg:CE} indicates good agreement between the two models for all transport coefficients, except the Spitzer correction to the parallel resistivity. 

By rotating coordinates to the current-aligned frame of reference, the resistivity coefficients from Eq.~(\ref{eq:eta_CE}) can be cast in the form
\begin{subequations}
\label{eq:eta_2_CE}
\begin{align}
\eta_J^{\textrm{CE}} &= \eta_\parallel^{\textrm{CE}} \cos^2\theta + \eta_\perp^{\textrm{CE}} \sin^2 \theta , \\ 
\eta_\times^{\textrm{CE}} &= (\eta_\perp^{\textrm{CE}} - \eta_\parallel^{\textrm{CE}}) \cos \theta \sin \theta , \\
\eta_\wedge^{\textrm{CE}} &= \eta_\wedge^{\textrm{CE}} .
\end{align}
\end{subequations} 
This form shows that the resistivity in the current-aligned direction, $\eta_J$, depends on the orientation of $\vc{J}$ with respect to $\vc{B}$. 
It also shows that there is a transverse resistivity coefficient in the magnetized regime, which for $\omega_{ce} \tau_e \gg 1$, asymptotes to $\eta_\times^{\textrm{CE}} \rightarrow 0.482 \eta_o \cos \theta \sin \theta$. 
This coefficient vanishes in the unmagnetized regime: $\eta_\times^{\textrm{CE}} \rightarrow 0$ for $\omega_{ce} \tau_e \ll 1$. 
The origin of these effects in the weakly magnetized regime is an asymmetry in the electron velocity distribution function either along or opposed to the magnetic field, in response to a macroscopic electric field. 

The conductivity coefficients to second order in the Chapman-Enskog expansion are~\cite{BraginskiiRPP1965,Ferziger1972} 
\begin{subequations}
\label{eq:sigma_CE}
\begin{align}
\sigma_\parallel^\textrm{CE} &= 1.931 \sigma_o \\
\sigma_\perp^\textrm{CE} &= \sigma_o \frac{(\omega_{ce} \tau_e)^2 + 1.802}{(\omega_{ce} \tau_e)^4 + 6.282 (\omega_{ce} \tau_e)^2 + 0.933} \\
\sigma_\wedge^\textrm{CE} &= - \frac{\sigma_o}{\omega_{ce} \tau_e} \frac{(\omega_{ce} \tau_e)^2 [(\omega_{ce} \tau_e)^2 + 4.382]}{(\omega_{ce} \tau_e)^4 + 6.282 (\omega_{ce} \tau_e)^2 + 0.933} .
\end{align}
\end{subequations}
Similarly to the resistivity coefficients, the main distinction between first and second order of the conductivity coefficients is the factor of 1.9 Spitzer correction to the parallel coefficient, which is independent of magnetic field strength. 
Both the perpendicular and cross coefficients are identical to the first order computation in the strongly magnetized regime $\omega_{ce} \tau_e \gg 1$, but are influenced by the Spitzer correction in the unmagnetized limit $\omega_{ce} \tau_e \ll 1$. 
These trends can be seen in Fig.~\ref{fg:CE}, along with the resistivity coefficients.

\section{Reduced formula for the friction force} 

This appendix derives reduced formula for the friction force in the limits that $\beta_e \ll 1$, as well as the limit that $\beta_e \rightarrow \infty$. 
For this development, it is useful to recall that the general dielectric response function from Eq.~(\ref{eq:full_epsilon}) can equivalently be expressed in the Gordeev integral representation~\cite{CavalierPOP2013,LafleurPPCF2019} 
\begin{equation}
\label{eq:gordeev}
\hat{\varepsilon}(\vc{k}, \omega) = 1 + \frac{1}{k^2 \lambda_{De}^2} \biggl[1 + iA\int_0^{\infty}dx e^{-B(1 - \cos x) - \frac{1}{2}Cx^2 + iAx} \biggr] ,
\end{equation} 
%in which $A = \omega/\omega_c = {\bf k}\cdot{\bf v}/\omega_c$, $B = k_{\perp}^2v_T^2/2\omega_c^2$, and $C = k_{\parallel}^2v_T^2/2\omega_c^2$, 
%Substituting equation~(\ref{eq:gordeev}) into (\ref{eq:force}), transforming to a spherical coordinate system such that, $d^3k = k^2\sin\theta^\prime dkd\theta^\prime d\phi^\prime$, $k_z = k\cos\theta^\prime$, $k_x = k\sin\theta^\prime \cos\phi^\prime$, $k_y = k\sin\theta^\prime \sin\phi^\prime $, $k_{\perp}^2 = k_x^2 + k_y^2 = k^2\sin^2\theta^\prime$, and simplifying, we obtain  
and the friction force components as
\begin{equation}
\label{eq:Fnorm_z}
\bar{F}_z = -\frac{6}{\pi^2} \int_0^{\pi/2}d\theta^\prime \sin\theta^\prime \cos\theta^\prime \int_0^{\pi}d\phi^\prime P(\theta^\prime, \phi^\prime) ,
\end{equation} 
and
\begin{equation}
\label{eq:Fnorm_x}
\bar{F}_x = -\frac{6}{\pi^2} \int_0^{\pi/2}d\theta^\prime \sin^2\theta^\prime  \int_0^{\pi} d\phi^\prime \cos\phi^\prime  P(\theta^\prime, \phi^\prime) .
\end{equation} 
Here,
\begin{equation}
\label{eq:P}
P(\theta^\prime, \phi^\prime) \equiv \int_0^{\bar{k}_\mx} d\bar{k} \frac{\bar{k}^3 G_i}{G_r^2 + G_i^2} ,
\end{equation}
\begin{equation}
G_r = 1+ \bar{k}^2 - A\int_0^{\infty}dx\sin(Ax)e^{-B(1 - \cos x) - \frac{1}{2}Cx^2},
\end{equation} 
\begin{equation}
\label{eq:Gi}
G_i = A\int_0^{\infty}dx\cos(Ax)e^{-B(1 - \cos x) - \frac{1}{2}Cx^2},
\end{equation} 
\begin{align}
\label{eq:A_def}
A &= \frac{\omega}{\omega_{ce}} = \frac{\vc{k} \cdot \vc{v}}{\omega_{ce}} \\ \nonumber
&= \frac{\sqrt{2} M \bar{k}}{\beta_e}\left(\sin\theta^\prime \cos\phi^\prime \sin\theta + \cos\theta^\prime \cos\theta \right),
\end{align} 
\begin{equation}
B = \frac{k_{\perp}^2v_{Te}^2}{2\omega_{ce}^2} = \frac{\bar{k}^2\sin^2\theta^\prime}{\beta_e^2},
\end{equation}
\begin{equation}
C = \frac{k_{\parallel}^2v_{Te}^2}{2\omega_{ce}^2}= \frac{\bar{k}^2\cos^2\theta^\prime}{\beta_e^2} ,
\end{equation} 
and the dimensionless quantities are $M=v/v_{Te}$, $\bar{k} \equiv k \lambda_{De}$, and $\bar{F} \equiv F a_e /(k_\bb T_e \Gamma_e^2)$ for each component of $F$. 

Conductivity is concerned with the limit that the ion flow is asymptotically small compared to the electron thermal speed: $M \ll 1$. 
In this limit, $A \ll 1$, so $G_r = 1 + \bar{k}^2 + \mathcal{O}(A^2)$, $G_i \sim \min \lbrace \mathcal{O}(A) \rbrace$, and 
\begin{equation}
\label{eq:P_sa}
P(\theta^\prime, \phi^\prime) = \int_0^{\bar{k}_\mx} d\bar{k} \frac{G_i \bar{k}^3}{(1+ \bar{k}^2)^2} + \mathcal{O}(A^4) .
\end{equation}
An important point with regard to the strongly magnetized limit is that, although $A\ll 1$, the $\cos(Ax)$ term in Eq.~(\ref{eq:Gi}) cannot be taken to be 1 because $x$ gets very large in the strongly magnetized regime. 
If one were to assume that $\cos (Ax)  =1$, the result diverges. 
In fact, this is the source of the nonlinearity of $\vc{F}$ with respect to $M$ in the strongly magnetized regime. 
The complexity of the integrand prevents any analytic simplification of Eq.~(\ref{eq:P_sa}) in the general case, but it can be simplified in both the weak and strong magnetization limits. 

Taking these limits, the next two subsections will show that in both the large and small $\beta_e$ limits, $G_i$ is independent of $\bar{k}$ to leading order. 
In either limit, the $\bar{k}$ integral in Eq.~(\ref{eq:P_sa}) gives the Coulomb logarithm 
\begin{eqnarray}
\label{eq:ln_lambda}
 \int_0^{\bar{k}_\mx} d\bar{k} \frac{\bar{k}^3}{(1+ \bar{k}^2)^2} &=& \ln (\sqrt{1 + \bar{k}_\mx^2}) - \frac{\bar{k}_\mx^2}{2(1+\bar{k}_\mx^2)} \nonumber \\
 &\approx& \ln \Lambda
\end{eqnarray}
where $\ln \Lambda = \ln \bar{k}_\mx$. 
One implication is that the friction force is proportional to the Coulomb logarithm in both the large and small $\beta_e$ limits, but not in general. 

%%%%%%%%%%
\subsection{Weakly magnetized limit~\label{ap:r_wm} }

In the weakly magnetized limit the gyromotion component of the particle trajectory, which is described by the $\cos x$ term in Eq.~(\ref{eq:Gi}), is small. 
This limit can be obtained from an expansion in small $x$: $1-\cos x = \frac{1}{2} x^2 - \frac{1}{24} x^4 + \ldots$.
Note that taking the $x\ll 1$ limit of Eq.~(\ref{eq:gordeev}) to first order results in Eq.~(\ref{eq:epsilon_o}). 
Here, we will carry out the expansion to second order, which for Eq.~(\ref{eq:Gi}) is
\begin{equation}
G_i = \int_0^\infty dy\, \cos(y) e^{-ay^2} \bigl(1 + \frac{B}{24 A^4} y^4 \bigr)
\end{equation}
where $y=Ax$ and $a \equiv (B+C)/(2A^2) = 1/[2M(\sin\theta^\prime \cos \phi^\prime \sin \theta + \cos \theta^\prime \cos \theta)]^2$. 
Integrating and taking the $a \gg 1$ limit (due to $M \ll 1$) provides 
\begin{equation}
G_i = \sqrt{\pi}M (\sin \theta^\prime \cos \phi^\prime \sin \theta + \cos \theta^\prime \cos \theta) \bigl(1 + \frac{\beta^2 \sin^2 \theta^\prime}{8\bar{k}^2}\bigr) .
\end{equation}
Making use of Eq.~(\ref{eq:ln_lambda}), Eq.~(\ref{eq:P_sa}) then reduces to 
\begin{equation}
P = \sqrt{\pi}M (\sin \theta^\prime \cos \phi^\prime \sin \theta + \cos \theta^\prime \cos \theta) \bigl(\ln \Lambda + \frac{\beta_e^2 \sin^2 \theta^\prime}{16}\bigr),
\end{equation}
showing that the first correction due to a finite magnetic field is independent of the Coulomb logarithm. 

The forces from Eqs.~(\ref{eq:Fnorm_z}) and (\ref{eq:Fnorm_x}) can then be evaluated analytically, resulting in
\begin{equation}
\label{eq:Fz_wb}
\bar{F}_z^{\beta_o}  = - \frac{2}{\sqrt{\pi}} M \cos \theta \bigl( \ln \Lambda + \frac{\beta_e^2}{40} \bigr) 
\end{equation}
and 
\begin{equation}
\label{eq:Fx}
\bar{F}_x^{\beta_o}  = - \frac{2}{\sqrt{\pi}} M \sin \theta \bigl( \ln \Lambda + \frac{\beta_e^2}{20} \bigr).
\end{equation} 
With these, the stopping power to second order is
\begin{equation}
\label{eq:Fv_so}
\bar{F}_v^{\beta_o}  = - \frac{2M}{\sqrt{\pi}} \biggl[ \ln \Lambda + \frac{\beta_e^2}{40}(\sin^2\theta + 1) \biggr]
\end{equation}
and the transverse force is 
\begin{equation}
\label{eq:Fx_so}
\bar{F}_\times^{\beta_o} = - \frac{M \sin \theta \cos \theta}{20 \sqrt{\pi}} \beta_e^2 .
\end{equation} 
As expected, Eq.~(\ref{eq:Fv_so}) shows that the first correction to the stopping power due to magnetization is small, since $\ln \Lambda \gg 1$ and $\beta_e \ll 1$ in this limit. 
The more meaningful result is Eq.~(\ref{eq:Fx_so}), which shows the first non-zero component of the transverse force. 
Although this is small in comparison to the stopping power in the weakly magnetized regime, it acts in a different direction and may become important after sufficient time has elapsed.  

%%%%%%%%%%
\subsection{$\beta_e \rightarrow \infty$ limit \label{app:beta_inf}} 

In the strongly magnetized limit ($\beta_e \rightarrow \infty$), the gyroradius of particles becomes small in comparison to all other length scales of relevance. 
Mathematically, this limit can be obtained by expanding for large $x$: $-B(1-\cos x) - \frac{1}{2} Cx^2 + iAx \approx -\frac{1}{2} C x^2 + iAx$. 
Note that applying this limit to Eq.~(\ref{eq:gordeev}) results in Eq.~(\ref{eq:epsilon_inf}). 
Here, we apply the expansion to first order for the forces. 
In this limit, Eq.~(\ref{eq:Gi}) reduces to
\begin{equation} 
G_i \approx \sqrt{\pi} \frac{A}{\sqrt{2C}} e^{-A^2/2C}
\end{equation}
where $A/\sqrt{2C} = M (\tan \theta^\prime \cos \phi^\prime \sin \theta + \cos \theta)$. 
Since $G_i$ is independent of $\bar{k}$ in this limit, Eq.~(\ref{eq:ln_lambda}) provides the solution to the $\bar{k}$ integral of Eq.~(\ref{eq:P_sa}), showing that the forces are proportional to the Coulomb logarithm 
\begin{equation}
P \approx \sqrt{\pi}M \ln \Lambda g(\theta^\prime, \phi^\prime, \theta) e^{-M^2 g^2(\theta^\prime, \phi^\prime, \theta)}
\end{equation}
where $g(\theta^\prime, \phi^\prime, \theta) \equiv \tan \theta^\prime \cos \phi^\prime \sin \theta + \cos \theta$. 
Applying this to Eqs.~(\ref{eq:Fnorm_z}) and (\ref{eq:Fnorm_x}), making the variable substitutions $y=\tan \theta^\prime$ and $w= \cos \phi^\prime$, then replacing the $y$ integral with $z \equiv yw$, and making use of the result 
\begin{equation}
\int_0^1 dw \frac{w^2}{\sqrt{1-w^2} (w^2 + z^2)^{3/2}} = \frac{\pi}{4} \frac{1}{|z|(1+z^2)^{3/2}} 
\end{equation} 
provides 
\begin{equation}
\bar{F}_z = - \frac{3 M \ln \Lambda}{2\sqrt{\pi}} \int_{-\infty}^\infty dz \frac{z \sin \theta + \cos \theta}{(1+z^2)^{3/2}} e^{-M^2 (z\sin \theta + \cos \theta)^2}
\end{equation}
and 
\begin{equation}
\label{eq:Fx_mid}
\bar{F}_x = - \frac{3 M \ln \Lambda}{2\sqrt{\pi}} \int_{-\infty}^\infty dz \frac{z(z \sin \theta + \cos \theta)}{(1+z^2)^{3/2}} e^{-M^2 (z\sin \theta + \cos \theta)^2}.
\end{equation}

Finally, we take the $M\ll 1$ limit. 
Nominally, this is expected to render the exponential term to be approximately 1. 
Indeed, for $\bar{F}_z$, doing so leads to 
\begin{equation}
\bar{F}_z^{\beta_\infty} = - \frac{3 \ln \Lambda}{\sqrt{\pi}} M \cos \theta . 
\end{equation}
However, the exponential term in $\bar{F}_x$ is essential because it resolves an otherwise divergent integral. 
This is the source of the nonlinearity of the friction force with respect to speed. 
The term proportional to $\cos \theta$ in Eq.~(\ref{eq:Fx_mid}) is very small compared to the first term because it is approximately an odd function of $z$ and asymptotes identically to zero in the limit $M \ll 1$. 
When $M \ll 1$, the exponential term acts to truncate the first term of the integrand at the value at which its argument is 1: $M^2(\cos \theta + z_c \sin \theta)^2 \approx 1 \Rightarrow z_c \approx  \pm 1/(M \sin \theta)$. 
With this
\begin{align}
\int_{-z_c}^{z_c} \frac{z^2}{(1+z^2)^{3/2}} &= 2 \textrm{arcsinh}(z_c) - \frac{2 z_c}{\sqrt{1+z_c^2}} \\ \nonumber
& \approx 2 \frac{z_c}{|z_c|} \ln |z_c| = -2 \frac{\sin \theta}{|\sin \theta|} \ln |M \sin \theta|,
\end{align}
so to first order in $M \ll 1$ 
\begin{equation}
\bar{F}_x^{\beta_\infty} = \frac{3 \ln \Lambda}{\sqrt{\pi}} M_\perp\, \ln M_\perp ,
\end{equation}
where $M_\perp = M |\sin \theta|$. 
Thus, the friction force is not predicted to be linearly proportional to $M$ in the $\beta_e \rightarrow \infty$ limit. 
In fact, the $M \sin\theta \ln (M \sin\theta)$ component shows that the friction force does not have a Taylor series in the $\beta_e \rightarrow \infty$ limit.
% calling into question the validity of a linear regime of transport. 
With these, the stopping power can be expressed as 
\begin{equation}
\label{eq:sp}
\bar{F}_v^{\beta_\infty} = - \frac{3 \ln \Lambda}{\sqrt{\pi}} M [ \cos^2 \theta - \sin^2 \theta \ln M_\perp] 
\end{equation}
and the transverse force as
\begin{equation}
\label{eq:Ftimes_binf}
\bar{F}_\times^{\beta_\infty} = \frac{3 \ln \Lambda}{\sqrt{\pi}} M \sin \theta\, \cos \theta [1 + \ln M_\perp] .
\end{equation} 
The logarithmic term is expected to be much larger than the first term in Eq.~(\ref{eq:Ftimes_binf}). 
Note that since $\ln M_\perp < 0$, the second term of Eq.~(\ref{eq:sp}) is always positive.

%%%%%%%%%%%%
\begin{acknowledgments}
This material is based upon work supported by the U.S. Department of Energy, Office of Fusion Energy Sciences under Award Number DE-SC0016159. 
\end{acknowledgments}

\bibliography{refs.bib}

\begin{thebibliography}{58}%
\makeatletter
\providecommand \@ifxundefined [1]{%
 \@ifx{#1\undefined}
}%
\providecommand \@ifnum [1]{%
 \ifnum #1\expandafter \@firstoftwo
 \else \expandafter \@secondoftwo
 \fi
}%
\providecommand \@ifx [1]{%
 \ifx #1\expandafter \@firstoftwo
 \else \expandafter \@secondoftwo
 \fi
}%
\providecommand \natexlab [1]{#1}%
\providecommand \enquote  [1]{``#1''}%
\providecommand \bibnamefont  [1]{#1}%
\providecommand \bibfnamefont [1]{#1}%
\providecommand \citenamefont [1]{#1}%
\providecommand \href@noop [0]{\@secondoftwo}%
\providecommand \href [0]{\begingroup \@sanitize@url \@href}%
\providecommand \@href[1]{\@@startlink{#1}\@@href}%
\providecommand \@@href[1]{\endgroup#1\@@endlink}%
\providecommand \@sanitize@url [0]{\catcode `\\12\catcode `\$12\catcode
  `\&12\catcode `\#12\catcode `\^12\catcode `\_12\catcode `\%12\relax}%
\providecommand \@@startlink[1]{}%
\providecommand \@@endlink[0]{}%
\providecommand \url  [0]{\begingroup\@sanitize@url \@url }%
\providecommand \@url [1]{\endgroup\@href {#1}{\urlprefix }}%
\providecommand \urlprefix  [0]{URL }%
\providecommand \Eprint [0]{\href }%
\providecommand \doibase [0]{http://dx.doi.org/}%
\providecommand \selectlanguage [0]{\@gobble}%
\providecommand \bibinfo  [0]{\@secondoftwo}%
\providecommand \bibfield  [0]{\@secondoftwo}%
\providecommand \translation [1]{[#1]}%
\providecommand \BibitemOpen [0]{}%
\providecommand \bibitemStop [0]{}%
\providecommand \bibitemNoStop [0]{.\EOS\space}%
\providecommand \EOS [0]{\spacefactor3000\relax}%
\providecommand \BibitemShut  [1]{\csname bibitem#1\endcsname}%
\let\auto@bib@innerbib\@empty
%</preamble>
\bibitem [{\citenamefont {{Braginskii}}(1965)}]{BraginskiiRPP1965}%
  \BibitemOpen
  \bibfield  {author} {\bibinfo {author} {\bibfnamefont {S.~I.}\ \bibnamefont
  {{Braginskii}}},\ }\href@noop {} {\bibfield  {journal} {\bibinfo  {journal}
  {Rev. Plasma Phys.}\ }\textbf {\bibinfo {volume} {vol. 1}} (\bibinfo {year}
  {1965})}\BibitemShut {NoStop}%
\bibitem [{\citenamefont {Kulsrud}\ and\ \citenamefont
  {Zweibel}(2008)}]{Kulsrud_2008}%
  \BibitemOpen
  \bibfield  {author} {\bibinfo {author} {\bibfnamefont {R.~M.}\ \bibnamefont
  {Kulsrud}}\ and\ \bibinfo {author} {\bibfnamefont {E.~G.}\ \bibnamefont
  {Zweibel}},\ }\href {\doibase 10.1088/0034-4885/71/4/046901} {\bibfield
  {journal} {\bibinfo  {journal} {Reports on Progress in Physics}\ }\textbf
  {\bibinfo {volume} {71}},\ \bibinfo {pages} {046901} (\bibinfo {year}
  {2008})}\BibitemShut {NoStop}%
\bibitem [{\citenamefont {Fox}\ \emph {et~al.}(2018)\citenamefont {Fox},
  \citenamefont {Matteucci}, \citenamefont {Moissard}, \citenamefont
  {Schaeffer}, \citenamefont {Bhattacharjee}, \citenamefont {Germaschewski},\
  and\ \citenamefont {Hu}}]{FoxPOP2018}%
  \BibitemOpen
  \bibfield  {author} {\bibinfo {author} {\bibfnamefont {W.}~\bibnamefont
  {Fox}}, \bibinfo {author} {\bibfnamefont {J.}~\bibnamefont {Matteucci}},
  \bibinfo {author} {\bibfnamefont {C.}~\bibnamefont {Moissard}}, \bibinfo
  {author} {\bibfnamefont {D.~B.}\ \bibnamefont {Schaeffer}}, \bibinfo {author}
  {\bibfnamefont {A.}~\bibnamefont {Bhattacharjee}}, \bibinfo {author}
  {\bibfnamefont {K.}~\bibnamefont {Germaschewski}}, \ and\ \bibinfo {author}
  {\bibfnamefont {S.~X.}\ \bibnamefont {Hu}},\ }\href {\doibase
  10.1063/1.5050813} {\bibfield  {journal} {\bibinfo  {journal} {Physics of
  Plasmas}\ }\textbf {\bibinfo {volume} {25}},\ \bibinfo {pages} {102106}
  (\bibinfo {year} {2018})},\ \Eprint
  {http://arxiv.org/abs/https://doi.org/10.1063/1.5050813}
  {https://doi.org/10.1063/1.5050813} \BibitemShut {NoStop}%
\bibitem [{\citenamefont {Gurnett}\ and\ \citenamefont
  {Bhattacharjee}(2005)}]{gurnett_bhattacharjee_2005}%
  \BibitemOpen
  \bibfield  {author} {\bibinfo {author} {\bibfnamefont {D.~A.}\ \bibnamefont
  {Gurnett}}\ and\ \bibinfo {author} {\bibfnamefont {A.}~\bibnamefont
  {Bhattacharjee}},\ }\href {\doibase 10.1017/CBO9780511809125} {\emph
  {\bibinfo {title} {Introduction to Plasma Physics: With Space and Laboratory
  Applications}}}\ (\bibinfo  {publisher} {Cambridge University Press},\
  \bibinfo {year} {2005})\BibitemShut {NoStop}%
\bibitem [{\citenamefont {Rax}\ \emph {et~al.}(2019)\citenamefont {Rax},
  \citenamefont {Kolmes}, \citenamefont {Ochs}, \citenamefont {Fisch},\ and\
  \citenamefont {Gueroult}}]{RaxPOP2019}%
  \BibitemOpen
  \bibfield  {author} {\bibinfo {author} {\bibfnamefont {J.~M.}\ \bibnamefont
  {Rax}}, \bibinfo {author} {\bibfnamefont {E.~J.}\ \bibnamefont {Kolmes}},
  \bibinfo {author} {\bibfnamefont {I.~E.}\ \bibnamefont {Ochs}}, \bibinfo
  {author} {\bibfnamefont {N.~J.}\ \bibnamefont {Fisch}}, \ and\ \bibinfo
  {author} {\bibfnamefont {R.}~\bibnamefont {Gueroult}},\ }\href {\doibase
  10.1063/1.5064520} {\bibfield  {journal} {\bibinfo  {journal} {Physics of
  Plasmas}\ }\textbf {\bibinfo {volume} {26}},\ \bibinfo {pages} {012303}
  (\bibinfo {year} {2019})},\ \Eprint
  {http://arxiv.org/abs/https://doi.org/10.1063/1.5064520}
  {https://doi.org/10.1063/1.5064520} \BibitemShut {NoStop}%
\bibitem [{\citenamefont {Yamada}, \citenamefont {Kulsrud},\ and\ \citenamefont
  {Ji}(2010)}]{YamadaRMP2010}%
  \BibitemOpen
  \bibfield  {author} {\bibinfo {author} {\bibfnamefont {M.}~\bibnamefont
  {Yamada}}, \bibinfo {author} {\bibfnamefont {R.}~\bibnamefont {Kulsrud}}, \
  and\ \bibinfo {author} {\bibfnamefont {H.}~\bibnamefont {Ji}},\ }\href
  {\doibase 10.1103/RevModPhys.82.603} {\bibfield  {journal} {\bibinfo
  {journal} {Rev. Mod. Phys.}\ }\textbf {\bibinfo {volume} {82}},\ \bibinfo
  {pages} {603} (\bibinfo {year} {2010})}\BibitemShut {NoStop}%
\bibitem [{\citenamefont {Plunian}\ and\ \citenamefont
  {Alboussière}(2021)}]{plunian_alboussiere_2021}%
  \BibitemOpen
  \bibfield  {author} {\bibinfo {author} {\bibfnamefont {F.}~\bibnamefont
  {Plunian}}\ and\ \bibinfo {author} {\bibfnamefont {T.}~\bibnamefont
  {Alboussière}},\ }\href {\doibase 10.1017/S0022377820001634} {\bibfield
  {journal} {\bibinfo  {journal} {Journal of Plasma Physics}\ }\textbf
  {\bibinfo {volume} {87}},\ \bibinfo {pages} {905870110} (\bibinfo {year}
  {2021})}\BibitemShut {NoStop}%
\bibitem [{\citenamefont {Wesson}(2011)}]{Wesson:1427009}%
  \BibitemOpen
  \bibfield  {author} {\bibinfo {author} {\bibfnamefont {J.}~\bibnamefont
  {Wesson}},\ }\href {https://cds.cern.ch/record/1427009} {\emph {\bibinfo
  {title} {{Tokamaks; 4th ed.}}}},\ International series of monographs on
  physics\ (\bibinfo  {publisher} {Oxford Univ. Press},\ \bibinfo {address}
  {Oxford},\ \bibinfo {year} {2011})\BibitemShut {NoStop}%
\bibitem [{\citenamefont {Chapman}\ and\ \citenamefont
  {Cowling}(1991)}]{Chapman:1991}%
  \BibitemOpen
  \bibfield  {author} {\bibinfo {author} {\bibfnamefont {S.}~\bibnamefont
  {Chapman}}\ and\ \bibinfo {author} {\bibfnamefont {T.~G.}\ \bibnamefont
  {Cowling}},\ }\href@noop {} {\emph {\bibinfo {title} {The Mathematical Theory
  of Non-Uniform Gases}}},\ \bibinfo {edition} {3rd}\ ed.\ (\bibinfo
  {publisher} {Cambridge University Press},\ \bibinfo {address} {Cambridge,
  UK},\ \bibinfo {year} {1991})\BibitemShut {NoStop}%
\bibitem [{\citenamefont {Landau}(1936)}]{LandauPZS1936}%
  \BibitemOpen
  \bibfield  {author} {\bibinfo {author} {\bibfnamefont {L.}~\bibnamefont
  {Landau}},\ }\href@noop {} {\bibfield  {journal} {\bibinfo  {journal} {Phys.
  Z. Sowjetunion}\ }\textbf {\bibinfo {volume} {10}},\ \bibinfo {pages} {154}
  (\bibinfo {year} {1936})}\BibitemShut {NoStop}%
\bibitem [{\citenamefont {Spitzer}\ and\ \citenamefont
  {H\"arm}(1953)}]{Spitzer:1953}%
  \BibitemOpen
  \bibfield  {author} {\bibinfo {author} {\bibfnamefont {L.}~\bibnamefont
  {Spitzer}}\ and\ \bibinfo {author} {\bibfnamefont {R.}~\bibnamefont
  {H\"arm}},\ }\href@noop {} {\bibfield  {journal} {\bibinfo  {journal} {Phys.
  Rev.}\ }\textbf {\bibinfo {volume} {89}},\ \bibinfo {pages} {977} (\bibinfo
  {year} {1953})}\BibitemShut {NoStop}%
\bibitem [{\citenamefont {{Book}}(1983)}]{NRL1983}%
  \BibitemOpen
  \bibfield  {author} {\bibinfo {author} {\bibfnamefont {D.~L.}\ \bibnamefont
  {{Book}}},\ }\href@noop {} {\enquote {\bibinfo {title} {{NRL (Naval Research
  Laboratory) plasma formulary, revised}},}\ }\bibinfo {howpublished} {Naval
  Research Lab. Report} (\bibinfo {year} {1983})\BibitemShut {NoStop}%
\bibitem [{\citenamefont {Ferziger}\ and\ \citenamefont
  {Kaper}(1972)}]{Ferziger1972}%
  \BibitemOpen
  \bibfield  {author} {\bibinfo {author} {\bibfnamefont {J.}~\bibnamefont
  {Ferziger}}\ and\ \bibinfo {author} {\bibfnamefont {H.}~\bibnamefont
  {Kaper}},\ }\href {https://books.google.com/books?id=qqm3AAAAIAAJ} {\emph
  {\bibinfo {title} {Mathematical Theory of Transport Processes in Gases}}}\
  (\bibinfo  {publisher} {North-Holland Publishing Company},\ \bibinfo {year}
  {1972})\BibitemShut {NoStop}%
\bibitem [{\citenamefont {Lenard}(1960)}]{LenardAP1960}%
  \BibitemOpen
  \bibfield  {author} {\bibinfo {author} {\bibfnamefont {A.}~\bibnamefont
  {Lenard}},\ }\href {\doibase https://doi.org/10.1016/0003-4916(60)90003-8}
  {\bibfield  {journal} {\bibinfo  {journal} {Annals of Physics}\ }\textbf
  {\bibinfo {volume} {10}},\ \bibinfo {pages} {390} (\bibinfo {year}
  {1960})}\BibitemShut {NoStop}%
\bibitem [{\citenamefont {Balescu}(1960)}]{BalescuPOP1960}%
  \BibitemOpen
  \bibfield  {author} {\bibinfo {author} {\bibfnamefont {R.}~\bibnamefont
  {Balescu}},\ }\href {\doibase 10.1063/1.1706002} {\bibfield  {journal}
  {\bibinfo  {journal} {The Physics of Fluids}\ }\textbf {\bibinfo {volume}
  {3}},\ \bibinfo {pages} {52} (\bibinfo {year} {1960})},\ \Eprint
  {http://arxiv.org/abs/https://aip.scitation.org/doi/pdf/10.1063/1.1706002}
  {https://aip.scitation.org/doi/pdf/10.1063/1.1706002} \BibitemShut {NoStop}%
\bibitem [{\citenamefont {Rostoker}(1960)}]{RostokerPF1960}%
  \BibitemOpen
  \bibfield  {author} {\bibinfo {author} {\bibfnamefont {N.}~\bibnamefont
  {Rostoker}},\ }\href {\doibase 10.1063/1.1706158} {\bibfield  {journal}
  {\bibinfo  {journal} {The Physics of Fluids}\ }\textbf {\bibinfo {volume}
  {3}},\ \bibinfo {pages} {922} (\bibinfo {year} {1960})},\ \Eprint
  {http://arxiv.org/abs/https://aip.scitation.org/doi/pdf/10.1063/1.1706158}
  {https://aip.scitation.org/doi/pdf/10.1063/1.1706158} \BibitemShut {NoStop}%
\bibitem [{\citenamefont {Nersisyan}, \citenamefont {Zwicknagel},\ and\
  \citenamefont {Toepffer}(2003)}]{NersisyanPRE2003}%
  \BibitemOpen
  \bibfield  {author} {\bibinfo {author} {\bibfnamefont {H.~B.}\ \bibnamefont
  {Nersisyan}}, \bibinfo {author} {\bibfnamefont {G.}~\bibnamefont
  {Zwicknagel}}, \ and\ \bibinfo {author} {\bibfnamefont {C.}~\bibnamefont
  {Toepffer}},\ }\href {\doibase 10.1103/PhysRevE.67.026411} {\bibfield
  {journal} {\bibinfo  {journal} {Phys. Rev. E}\ }\textbf {\bibinfo {volume}
  {67}},\ \bibinfo {pages} {026411} (\bibinfo {year} {2003})}\BibitemShut
  {NoStop}%
\bibitem [{\citenamefont {Nersisyan}, \citenamefont {Toepffer},\ and\
  \citenamefont {Zwicknagel}(2007)}]{Nersisyan:2007}%
  \BibitemOpen
  \bibfield  {author} {\bibinfo {author} {\bibfnamefont {H.}~\bibnamefont
  {Nersisyan}}, \bibinfo {author} {\bibfnamefont {C.}~\bibnamefont {Toepffer}},
  \ and\ \bibinfo {author} {\bibfnamefont {G.}~\bibnamefont {Zwicknagel}},\
  }\href@noop {} {\emph {\bibinfo {title} {Interactions Between Charged
  Particles in a Magnetic Field}}},\ \bibinfo {edition} {1st}\ ed.\ (\bibinfo
  {publisher} {Springer-Verlag},\ \bibinfo {address} {Berlin},\ \bibinfo {year}
  {2007})\BibitemShut {NoStop}%
\bibitem [{\citenamefont {ONeil}(1983)}]{OneilPF1983}%
  \BibitemOpen
  \bibfield  {author} {\bibinfo {author} {\bibfnamefont {T.~M.}\ \bibnamefont
  {ONeil}},\ }\href {\doibase 10.1063/1.864394} {\bibfield  {journal} {\bibinfo
   {journal} {The Physics of Fluids}\ }\textbf {\bibinfo {volume} {26}},\
  \bibinfo {pages} {2128} (\bibinfo {year} {1983})},\ \Eprint
  {http://arxiv.org/abs/https://aip.scitation.org/doi/pdf/10.1063/1.864394}
  {https://aip.scitation.org/doi/pdf/10.1063/1.864394} \BibitemShut {NoStop}%
\bibitem [{\citenamefont {Nersisyan}\ and\ \citenamefont
  {Zwicknagel}(2009)}]{NersisyanPRE2009}%
  \BibitemOpen
  \bibfield  {author} {\bibinfo {author} {\bibfnamefont {H.~B.}\ \bibnamefont
  {Nersisyan}}\ and\ \bibinfo {author} {\bibfnamefont {G.}~\bibnamefont
  {Zwicknagel}},\ }\href {\doibase 10.1103/PhysRevE.79.066405} {\bibfield
  {journal} {\bibinfo  {journal} {Phys. Rev. E}\ }\textbf {\bibinfo {volume}
  {79}},\ \bibinfo {pages} {066405} (\bibinfo {year} {2009})}\BibitemShut
  {NoStop}%
\bibitem [{\citenamefont {Jose}\ and\ \citenamefont
  {Baalrud}(2020)}]{JosePOP2020}%
  \BibitemOpen
  \bibfield  {author} {\bibinfo {author} {\bibfnamefont {L.}~\bibnamefont
  {Jose}}\ and\ \bibinfo {author} {\bibfnamefont {S.~D.}\ \bibnamefont
  {Baalrud}},\ }\href@noop {} {\bibfield  {journal} {\bibinfo  {journal} {Phys.
  Plasmas}\ }\textbf {\bibinfo {volume} {27}},\ \bibinfo {pages} {112101}
  (\bibinfo {year} {2020})}\BibitemShut {NoStop}%
\bibitem [{\citenamefont {Montgomery}, \citenamefont {Joyce},\ and\
  \citenamefont {Turner}(1974)}]{MontgomeryPF1974}%
  \BibitemOpen
  \bibfield  {author} {\bibinfo {author} {\bibfnamefont {D.}~\bibnamefont
  {Montgomery}}, \bibinfo {author} {\bibfnamefont {G.}~\bibnamefont {Joyce}}, \
  and\ \bibinfo {author} {\bibfnamefont {L.}~\bibnamefont {Turner}},\
  }\href@noop {} {\bibfield  {journal} {\bibinfo  {journal} {Phys. Fluids}\
  }\textbf {\bibinfo {volume} {17}},\ \bibinfo {pages} {2201} (\bibinfo {year}
  {1974})}\BibitemShut {NoStop}%
\bibitem [{\citenamefont {Ware}(1989)}]{WarePRL1989}%
  \BibitemOpen
  \bibfield  {author} {\bibinfo {author} {\bibfnamefont {A.~A.}\ \bibnamefont
  {Ware}},\ }\href {\doibase 10.1103/PhysRevLett.62.51} {\bibfield  {journal}
  {\bibinfo  {journal} {Phys. Rev. Lett.}\ }\textbf {\bibinfo {volume} {62}},\
  \bibinfo {pages} {51} (\bibinfo {year} {1989})}\BibitemShut {NoStop}%
\bibitem [{\citenamefont {Cohen}, \citenamefont {Sarid},\ and\ \citenamefont
  {Gedalin}(2019)}]{CohenPOP2019}%
  \BibitemOpen
  \bibfield  {author} {\bibinfo {author} {\bibfnamefont {S.}~\bibnamefont
  {Cohen}}, \bibinfo {author} {\bibfnamefont {E.}~\bibnamefont {Sarid}}, \ and\
  \bibinfo {author} {\bibfnamefont {M.}~\bibnamefont {Gedalin}},\ }\href
  {\doibase 10.1063/1.5109965} {\bibfield  {journal} {\bibinfo  {journal}
  {Physics of Plasmas}\ }\textbf {\bibinfo {volume} {26}},\ \bibinfo {pages}
  {082105} (\bibinfo {year} {2019})},\ \Eprint
  {http://arxiv.org/abs/https://doi.org/10.1063/1.5109965}
  {https://doi.org/10.1063/1.5109965} \BibitemShut {NoStop}%
\bibitem [{\citenamefont {Dubin}(2014)}]{DubinPOP2014}%
  \BibitemOpen
  \bibfield  {author} {\bibinfo {author} {\bibfnamefont {D.~H.~E.}\
  \bibnamefont {Dubin}},\ }\href {\doibase 10.1063/1.4876749} {\bibfield
  {journal} {\bibinfo  {journal} {Physics of Plasmas}\ }\textbf {\bibinfo
  {volume} {21}},\ \bibinfo {pages} {052108} (\bibinfo {year} {2014})},\
  \Eprint {http://arxiv.org/abs/https://doi.org/10.1063/1.4876749}
  {https://doi.org/10.1063/1.4876749} \BibitemShut {NoStop}%
\bibitem [{\citenamefont {Grad}(1958)}]{Grad1958}%
  \BibitemOpen
  \bibfield  {author} {\bibinfo {author} {\bibfnamefont {H.}~\bibnamefont
  {Grad}},\ }\href@noop {} {\emph {\bibinfo {title} {Principles of the kinetic
  theory of gases}}}\ (\bibinfo  {publisher} {Springer},\ \bibinfo {year}
  {1958})\BibitemShut {NoStop}%
\bibitem [{\citenamefont {Lafleur}\ and\ \citenamefont
  {Baalrud}(2019)}]{LafleurPPCF2019}%
  \BibitemOpen
  \bibfield  {author} {\bibinfo {author} {\bibfnamefont {T.}~\bibnamefont
  {Lafleur}}\ and\ \bibinfo {author} {\bibfnamefont {S.~D.}\ \bibnamefont
  {Baalrud}},\ }\href@noop {} {\bibfield  {journal} {\bibinfo  {journal}
  {Plasma Phys. Control. Fusion}\ }\textbf {\bibinfo {volume} {61}},\ \bibinfo
  {pages} {125004} (\bibinfo {year} {2019})}\BibitemShut {NoStop}%
\bibitem [{\citenamefont {Lafleur}\ and\ \citenamefont
  {Baalrud}(2020)}]{LafleurPPCF2020}%
  \BibitemOpen
  \bibfield  {author} {\bibinfo {author} {\bibfnamefont {T.}~\bibnamefont
  {Lafleur}}\ and\ \bibinfo {author} {\bibfnamefont {S.~D.}\ \bibnamefont
  {Baalrud}},\ }\href {\doibase 10.1088/1361-6587/ab9bea} {\bibfield  {journal}
  {\bibinfo  {journal} {Plasma Physics and Controlled Fusion}\ }\textbf
  {\bibinfo {volume} {62}},\ \bibinfo {pages} {095003} (\bibinfo {year}
  {2020})}\BibitemShut {NoStop}%
\bibitem [{\citenamefont {Bernstein}\ \emph {et~al.}(2020)\citenamefont
  {Bernstein}, \citenamefont {Lafleur}, \citenamefont {Daligault},\ and\
  \citenamefont {Baalrud}}]{BernsteinPRE2020}%
  \BibitemOpen
  \bibfield  {author} {\bibinfo {author} {\bibfnamefont {D.~J.}\ \bibnamefont
  {Bernstein}}, \bibinfo {author} {\bibfnamefont {T.}~\bibnamefont {Lafleur}},
  \bibinfo {author} {\bibfnamefont {J.}~\bibnamefont {Daligault}}, \ and\
  \bibinfo {author} {\bibfnamefont {S.~D.}\ \bibnamefont {Baalrud}},\ }\href
  {\doibase 10.1103/PhysRevE.102.041201} {\bibfield  {journal} {\bibinfo
  {journal} {Phys. Rev. E}\ }\textbf {\bibinfo {volume} {102}},\ \bibinfo
  {pages} {041201} (\bibinfo {year} {2020})}\BibitemShut {NoStop}%
\bibitem [{\citenamefont {Cereceda}, \citenamefont {de~Peretti},\ and\
  \citenamefont {Deutsch}(2005)}]{CerecedaPOP2005}%
  \BibitemOpen
  \bibfield  {author} {\bibinfo {author} {\bibfnamefont {C.}~\bibnamefont
  {Cereceda}}, \bibinfo {author} {\bibfnamefont {M.}~\bibnamefont
  {de~Peretti}}, \ and\ \bibinfo {author} {\bibfnamefont {C.}~\bibnamefont
  {Deutsch}},\ }\href {\doibase 10.1063/1.1848545} {\bibfield  {journal}
  {\bibinfo  {journal} {Physics of Plasmas}\ }\textbf {\bibinfo {volume}
  {12}},\ \bibinfo {pages} {022102} (\bibinfo {year} {2005})},\ \Eprint
  {http://arxiv.org/abs/https://doi.org/10.1063/1.1848545}
  {https://doi.org/10.1063/1.1848545} \BibitemShut {NoStop}%
\bibitem [{\citenamefont {Baalrud}\ and\ \citenamefont
  {Daligault}(2017)}]{BaalrudPRE2017}%
  \BibitemOpen
  \bibfield  {author} {\bibinfo {author} {\bibfnamefont {S.~D.}\ \bibnamefont
  {Baalrud}}\ and\ \bibinfo {author} {\bibfnamefont {J.}~\bibnamefont
  {Daligault}},\ }\href@noop {} {\bibfield  {journal} {\bibinfo  {journal}
  {Phys. Rev. E}\ }\textbf {\bibinfo {volume} {96}} (\bibinfo {year}
  {2017})}\BibitemShut {NoStop}%
\bibitem [{\citenamefont {Dubin}(1998)}]{DubinPOP1998}%
  \BibitemOpen
  \bibfield  {author} {\bibinfo {author} {\bibfnamefont {D.~H.~E.}\
  \bibnamefont {Dubin}},\ }\href {\doibase 10.1063/1.872837} {\bibfield
  {journal} {\bibinfo  {journal} {Physics of Plasmas}\ }\textbf {\bibinfo
  {volume} {5}},\ \bibinfo {pages} {1688} (\bibinfo {year} {1998})},\ \Eprint
  {http://arxiv.org/abs/https://doi.org/10.1063/1.872837}
  {https://doi.org/10.1063/1.872837} \BibitemShut {NoStop}%
\bibitem [{\citenamefont {Fajans}\ and\ \citenamefont
  {Surko}(2020)}]{FajansPOP2020}%
  \BibitemOpen
  \bibfield  {author} {\bibinfo {author} {\bibfnamefont {J.}~\bibnamefont
  {Fajans}}\ and\ \bibinfo {author} {\bibfnamefont {C.~M.}\ \bibnamefont
  {Surko}},\ }\href {\doibase 10.1063/1.5131273} {\bibfield  {journal}
  {\bibinfo  {journal} {Physics of Plasmas}\ }\textbf {\bibinfo {volume}
  {27}},\ \bibinfo {pages} {030601} (\bibinfo {year} {2020})},\ \Eprint
  {http://arxiv.org/abs/https://doi.org/10.1063/1.5131273}
  {https://doi.org/10.1063/1.5131273} \BibitemShut {NoStop}%
\bibitem [{\citenamefont {Zhang}\ \emph {et~al.}(2008)\citenamefont {Zhang},
  \citenamefont {Fletcher}, \citenamefont {Rolston}, \citenamefont {Guzdar},\
  and\ \citenamefont {Swisdak}}]{ZhangPRL2008}%
  \BibitemOpen
  \bibfield  {author} {\bibinfo {author} {\bibfnamefont {X.~L.}\ \bibnamefont
  {Zhang}}, \bibinfo {author} {\bibfnamefont {R.~S.}\ \bibnamefont {Fletcher}},
  \bibinfo {author} {\bibfnamefont {S.~L.}\ \bibnamefont {Rolston}}, \bibinfo
  {author} {\bibfnamefont {P.~N.}\ \bibnamefont {Guzdar}}, \ and\ \bibinfo
  {author} {\bibfnamefont {M.}~\bibnamefont {Swisdak}},\ }\href@noop {}
  {\bibfield  {journal} {\bibinfo  {journal} {Phys. Rev. Lett.}\ }\textbf
  {\bibinfo {volume} {100}},\ \bibinfo {pages} {235002} (\bibinfo {year}
  {2008})}\BibitemShut {NoStop}%
\bibitem [{\citenamefont {Tiwari}\ and\ \citenamefont
  {Baalrud}(2018)}]{TiwariPOP2018}%
  \BibitemOpen
  \bibfield  {author} {\bibinfo {author} {\bibfnamefont {S.~K.}\ \bibnamefont
  {Tiwari}}\ and\ \bibinfo {author} {\bibfnamefont {S.~D.}\ \bibnamefont
  {Baalrud}},\ }\href {\doibase 10.1063/1.5013320} {\bibfield  {journal}
  {\bibinfo  {journal} {Physics of Plasmas}\ }\textbf {\bibinfo {volume}
  {25}},\ \bibinfo {pages} {013511} (\bibinfo {year} {2018})},\ \Eprint
  {http://arxiv.org/abs/https://doi.org/10.1063/1.5013320}
  {https://doi.org/10.1063/1.5013320} \BibitemShut {NoStop}%
\bibitem [{\citenamefont {Gorman}\ \emph {et~al.}(2021)\citenamefont {Gorman},
  \citenamefont {Warrens}, \citenamefont {Bradshaw},\ and\ \citenamefont
  {Killian}}]{GormanPRL2021}%
  \BibitemOpen
  \bibfield  {author} {\bibinfo {author} {\bibfnamefont {G.~M.}\ \bibnamefont
  {Gorman}}, \bibinfo {author} {\bibfnamefont {M.~K.}\ \bibnamefont {Warrens}},
  \bibinfo {author} {\bibfnamefont {S.~J.}\ \bibnamefont {Bradshaw}}, \ and\
  \bibinfo {author} {\bibfnamefont {T.~C.}\ \bibnamefont {Killian}},\ }\href
  {\doibase 10.1103/PhysRevLett.126.085002} {\bibfield  {journal} {\bibinfo
  {journal} {Phys. Rev. Lett.}\ }\textbf {\bibinfo {volume} {126}},\ \bibinfo
  {pages} {085002} (\bibinfo {year} {2021})}\BibitemShut {NoStop}%
\bibitem [{\citenamefont {Guthrie}\ and\ \citenamefont
  {Roberts}(2021)}]{GuthrieA2021}%
  \BibitemOpen
  \bibfield  {author} {\bibinfo {author} {\bibfnamefont {J.~M.}\ \bibnamefont
  {Guthrie}}\ and\ \bibinfo {author} {\bibfnamefont {J.~L.}\ \bibnamefont
  {Roberts}},\ }\href@noop {} {\enquote {\bibinfo {title} {Finite-amplitude rf
  heating rates for magnetized electrons in neutral plasma},}\ } (\bibinfo
  {year} {2021}),\ \Eprint {http://arxiv.org/abs/2104.02267} {arXiv:2104.02267
  [physics.plasm-ph]} \BibitemShut {NoStop}%
\bibitem [{\citenamefont {Thomas}, \citenamefont {Merlino},\ and\ \citenamefont
  {Rosenberg}(2012)}]{ThomasPPCF2012}%
  \BibitemOpen
  \bibfield  {author} {\bibinfo {author} {\bibfnamefont {E.}~\bibnamefont
  {Thomas}}, \bibinfo {author} {\bibfnamefont {R.~L.}\ \bibnamefont {Merlino}},
  \ and\ \bibinfo {author} {\bibfnamefont {M.}~\bibnamefont {Rosenberg}},\
  }\href@noop {} {\bibfield  {journal} {\bibinfo  {journal} {Plasma Phys.
  Control. Fusion}\ }\textbf {\bibinfo {volume} {54}},\ \bibinfo {pages}
  {124034} (\bibinfo {year} {2012})}\BibitemShut {NoStop}%
\bibitem [{\citenamefont {Bonitz}\ \emph {et~al.}(2012)\citenamefont {Bonitz},
  \citenamefont {Kalhlert}, \citenamefont {Ott},\ and\ \citenamefont
  {Laowen}}]{BonitzPSST2012}%
  \BibitemOpen
  \bibfield  {author} {\bibinfo {author} {\bibfnamefont {M.}~\bibnamefont
  {Bonitz}}, \bibinfo {author} {\bibfnamefont {H.}~\bibnamefont {Kalhlert}},
  \bibinfo {author} {\bibfnamefont {T.}~\bibnamefont {Ott}}, \ and\ \bibinfo
  {author} {\bibfnamefont {H.}~\bibnamefont {Laowen}},\ }\href@noop {}
  {\bibfield  {journal} {\bibinfo  {journal} {Plasma Sources Sci. Technol.}\
  }\textbf {\bibinfo {volume} {22}},\ \bibinfo {pages} {015007} (\bibinfo
  {year} {2012})}\BibitemShut {NoStop}%
\bibitem [{\citenamefont {Hartmann}\ \emph {et~al.}(2019)\citenamefont
  {Hartmann}, \citenamefont {Reyes}, \citenamefont {Kostadinova}, \citenamefont
  {Matthews}, \citenamefont {Hyde}, \citenamefont {Masheyeva}, \citenamefont
  {Dzhumagulova}, \citenamefont {Ramazanov}, \citenamefont {Ott}, \citenamefont
  {K\"ahlert}, \citenamefont {Bonitz}, \citenamefont {Korolov},\ and\
  \citenamefont {Donk\'o}}]{HartmannPRE2019}%
  \BibitemOpen
  \bibfield  {author} {\bibinfo {author} {\bibfnamefont {P.}~\bibnamefont
  {Hartmann}}, \bibinfo {author} {\bibfnamefont {J.~C.}\ \bibnamefont {Reyes}},
  \bibinfo {author} {\bibfnamefont {E.~G.}\ \bibnamefont {Kostadinova}},
  \bibinfo {author} {\bibfnamefont {L.~S.}\ \bibnamefont {Matthews}}, \bibinfo
  {author} {\bibfnamefont {T.~W.}\ \bibnamefont {Hyde}}, \bibinfo {author}
  {\bibfnamefont {R.~U.}\ \bibnamefont {Masheyeva}}, \bibinfo {author}
  {\bibfnamefont {K.~N.}\ \bibnamefont {Dzhumagulova}}, \bibinfo {author}
  {\bibfnamefont {T.~S.}\ \bibnamefont {Ramazanov}}, \bibinfo {author}
  {\bibfnamefont {T.}~\bibnamefont {Ott}}, \bibinfo {author} {\bibfnamefont
  {H.}~\bibnamefont {K\"ahlert}}, \bibinfo {author} {\bibfnamefont
  {M.}~\bibnamefont {Bonitz}}, \bibinfo {author} {\bibfnamefont
  {I.}~\bibnamefont {Korolov}}, \ and\ \bibinfo {author} {\bibfnamefont
  {Z.}~\bibnamefont {Donk\'o}},\ }\href {\doibase 10.1103/PhysRevE.99.013203}
  {\bibfield  {journal} {\bibinfo  {journal} {Phys. Rev. E}\ }\textbf {\bibinfo
  {volume} {99}},\ \bibinfo {pages} {013203} (\bibinfo {year}
  {2019})}\BibitemShut {NoStop}%
\bibitem [{\citenamefont {Gomez}\ \emph {et~al.}(2014)\citenamefont {Gomez},
  \citenamefont {Slutz}, \citenamefont {Sefkow}, \citenamefont {Sinars},
  \citenamefont {Hahn}, \citenamefont {Hansen}, \citenamefont {Harding},
  \citenamefont {Knapp}, \citenamefont {Schmit}, \citenamefont {Jennings},
  \citenamefont {Awe}, \citenamefont {Geissel}, \citenamefont {Rovang},
  \citenamefont {Chandler}, \citenamefont {Cooper}, \citenamefont {Cuneo},
  \citenamefont {Harvey-Thompson}, \citenamefont {Herrmann}, \citenamefont
  {Hess}, \citenamefont {Johns}, \citenamefont {Lamppa}, \citenamefont
  {Martin}, \citenamefont {McBride}, \citenamefont {Peterson}, \citenamefont
  {Porter}, \citenamefont {Robertson}, \citenamefont {Rochau}, \citenamefont
  {Ruiz}, \citenamefont {Savage}, \citenamefont {Smith}, \citenamefont
  {Stygar},\ and\ \citenamefont {Vesey}}]{GomezPRL2014}%
  \BibitemOpen
  \bibfield  {author} {\bibinfo {author} {\bibfnamefont {M.~R.}\ \bibnamefont
  {Gomez}}, \bibinfo {author} {\bibfnamefont {S.~A.}\ \bibnamefont {Slutz}},
  \bibinfo {author} {\bibfnamefont {A.~B.}\ \bibnamefont {Sefkow}}, \bibinfo
  {author} {\bibfnamefont {D.~B.}\ \bibnamefont {Sinars}}, \bibinfo {author}
  {\bibfnamefont {K.~D.}\ \bibnamefont {Hahn}}, \bibinfo {author}
  {\bibfnamefont {S.~B.}\ \bibnamefont {Hansen}}, \bibinfo {author}
  {\bibfnamefont {E.~C.}\ \bibnamefont {Harding}}, \bibinfo {author}
  {\bibfnamefont {P.~F.}\ \bibnamefont {Knapp}}, \bibinfo {author}
  {\bibfnamefont {P.~F.}\ \bibnamefont {Schmit}}, \bibinfo {author}
  {\bibfnamefont {C.~A.}\ \bibnamefont {Jennings}}, \bibinfo {author}
  {\bibfnamefont {T.~J.}\ \bibnamefont {Awe}}, \bibinfo {author} {\bibfnamefont
  {M.}~\bibnamefont {Geissel}}, \bibinfo {author} {\bibfnamefont {D.~C.}\
  \bibnamefont {Rovang}}, \bibinfo {author} {\bibfnamefont {G.~A.}\
  \bibnamefont {Chandler}}, \bibinfo {author} {\bibfnamefont {G.~W.}\
  \bibnamefont {Cooper}}, \bibinfo {author} {\bibfnamefont {M.~E.}\
  \bibnamefont {Cuneo}}, \bibinfo {author} {\bibfnamefont {A.~J.}\ \bibnamefont
  {Harvey-Thompson}}, \bibinfo {author} {\bibfnamefont {M.~C.}\ \bibnamefont
  {Herrmann}}, \bibinfo {author} {\bibfnamefont {M.~H.}\ \bibnamefont {Hess}},
  \bibinfo {author} {\bibfnamefont {O.}~\bibnamefont {Johns}}, \bibinfo
  {author} {\bibfnamefont {D.~C.}\ \bibnamefont {Lamppa}}, \bibinfo {author}
  {\bibfnamefont {M.~R.}\ \bibnamefont {Martin}}, \bibinfo {author}
  {\bibfnamefont {R.~D.}\ \bibnamefont {McBride}}, \bibinfo {author}
  {\bibfnamefont {K.~J.}\ \bibnamefont {Peterson}}, \bibinfo {author}
  {\bibfnamefont {J.~L.}\ \bibnamefont {Porter}}, \bibinfo {author}
  {\bibfnamefont {G.~K.}\ \bibnamefont {Robertson}}, \bibinfo {author}
  {\bibfnamefont {G.~A.}\ \bibnamefont {Rochau}}, \bibinfo {author}
  {\bibfnamefont {C.~L.}\ \bibnamefont {Ruiz}}, \bibinfo {author}
  {\bibfnamefont {M.~E.}\ \bibnamefont {Savage}}, \bibinfo {author}
  {\bibfnamefont {I.~C.}\ \bibnamefont {Smith}}, \bibinfo {author}
  {\bibfnamefont {W.~A.}\ \bibnamefont {Stygar}}, \ and\ \bibinfo {author}
  {\bibfnamefont {R.~A.}\ \bibnamefont {Vesey}},\ }\href {\doibase
  10.1103/PhysRevLett.113.155003} {\bibfield  {journal} {\bibinfo  {journal}
  {Phys. Rev. Lett.}\ }\textbf {\bibinfo {volume} {113}},\ \bibinfo {pages}
  {155003} (\bibinfo {year} {2014})}\BibitemShut {NoStop}%
\bibitem [{\citenamefont {Gomez}\ \emph {et~al.}(2020)\citenamefont {Gomez},
  \citenamefont {Slutz}, \citenamefont {Jennings}, \citenamefont {Ampleford},
  \citenamefont {Weis}, \citenamefont {Myers}, \citenamefont {Yager-Elorriaga},
  \citenamefont {Hahn}, \citenamefont {Hansen}, \citenamefont {Harding},
  \citenamefont {Harvey-Thompson}, \citenamefont {Lamppa}, \citenamefont
  {Mangan}, \citenamefont {Knapp}, \citenamefont {Awe}, \citenamefont
  {Chandler}, \citenamefont {Cooper}, \citenamefont {Fein}, \citenamefont
  {Geissel}, \citenamefont {Glinsky}, \citenamefont {Lewis}, \citenamefont
  {Ruiz}, \citenamefont {Ruiz}, \citenamefont {Savage}, \citenamefont {Schmit},
  \citenamefont {Smith}, \citenamefont {Styron}, \citenamefont {Porter},
  \citenamefont {Jones}, \citenamefont {Mattsson}, \citenamefont {Peterson},
  \citenamefont {Rochau},\ and\ \citenamefont {Sinars}}]{GomezPRL2020}%
  \BibitemOpen
  \bibfield  {author} {\bibinfo {author} {\bibfnamefont {M.~R.}\ \bibnamefont
  {Gomez}}, \bibinfo {author} {\bibfnamefont {S.~A.}\ \bibnamefont {Slutz}},
  \bibinfo {author} {\bibfnamefont {C.~A.}\ \bibnamefont {Jennings}}, \bibinfo
  {author} {\bibfnamefont {D.~J.}\ \bibnamefont {Ampleford}}, \bibinfo {author}
  {\bibfnamefont {M.~R.}\ \bibnamefont {Weis}}, \bibinfo {author}
  {\bibfnamefont {C.~E.}\ \bibnamefont {Myers}}, \bibinfo {author}
  {\bibfnamefont {D.~A.}\ \bibnamefont {Yager-Elorriaga}}, \bibinfo {author}
  {\bibfnamefont {K.~D.}\ \bibnamefont {Hahn}}, \bibinfo {author}
  {\bibfnamefont {S.~B.}\ \bibnamefont {Hansen}}, \bibinfo {author}
  {\bibfnamefont {E.~C.}\ \bibnamefont {Harding}}, \bibinfo {author}
  {\bibfnamefont {A.~J.}\ \bibnamefont {Harvey-Thompson}}, \bibinfo {author}
  {\bibfnamefont {D.~C.}\ \bibnamefont {Lamppa}}, \bibinfo {author}
  {\bibfnamefont {M.}~\bibnamefont {Mangan}}, \bibinfo {author} {\bibfnamefont
  {P.~F.}\ \bibnamefont {Knapp}}, \bibinfo {author} {\bibfnamefont {T.~J.}\
  \bibnamefont {Awe}}, \bibinfo {author} {\bibfnamefont {G.~A.}\ \bibnamefont
  {Chandler}}, \bibinfo {author} {\bibfnamefont {G.~W.}\ \bibnamefont
  {Cooper}}, \bibinfo {author} {\bibfnamefont {J.~R.}\ \bibnamefont {Fein}},
  \bibinfo {author} {\bibfnamefont {M.}~\bibnamefont {Geissel}}, \bibinfo
  {author} {\bibfnamefont {M.~E.}\ \bibnamefont {Glinsky}}, \bibinfo {author}
  {\bibfnamefont {W.~E.}\ \bibnamefont {Lewis}}, \bibinfo {author}
  {\bibfnamefont {C.~L.}\ \bibnamefont {Ruiz}}, \bibinfo {author}
  {\bibfnamefont {D.~E.}\ \bibnamefont {Ruiz}}, \bibinfo {author}
  {\bibfnamefont {M.~E.}\ \bibnamefont {Savage}}, \bibinfo {author}
  {\bibfnamefont {P.~F.}\ \bibnamefont {Schmit}}, \bibinfo {author}
  {\bibfnamefont {I.~C.}\ \bibnamefont {Smith}}, \bibinfo {author}
  {\bibfnamefont {J.~D.}\ \bibnamefont {Styron}}, \bibinfo {author}
  {\bibfnamefont {J.~L.}\ \bibnamefont {Porter}}, \bibinfo {author}
  {\bibfnamefont {B.}~\bibnamefont {Jones}}, \bibinfo {author} {\bibfnamefont
  {T.~R.}\ \bibnamefont {Mattsson}}, \bibinfo {author} {\bibfnamefont {K.~J.}\
  \bibnamefont {Peterson}}, \bibinfo {author} {\bibfnamefont {G.~A.}\
  \bibnamefont {Rochau}}, \ and\ \bibinfo {author} {\bibfnamefont {D.~B.}\
  \bibnamefont {Sinars}},\ }\href {\doibase 10.1103/PhysRevLett.125.155002}
  {\bibfield  {journal} {\bibinfo  {journal} {Phys. Rev. Lett.}\ }\textbf
  {\bibinfo {volume} {125}},\ \bibinfo {pages} {155002} (\bibinfo {year}
  {2020})}\BibitemShut {NoStop}%
\bibitem [{\citenamefont {Tatarakis}\ \emph {et~al.}(2002)\citenamefont
  {Tatarakis}, \citenamefont {Gopal}, \citenamefont {Watts}, \citenamefont
  {Beg}, \citenamefont {Dangor}, \citenamefont {Krushelnick}, \citenamefont
  {Wagner}, \citenamefont {Norreys}, \citenamefont {Clark}, \citenamefont
  {Zepf},\ and\ \citenamefont {Evans}}]{TatarakisPOP2002}%
  \BibitemOpen
  \bibfield  {author} {\bibinfo {author} {\bibfnamefont {M.}~\bibnamefont
  {Tatarakis}}, \bibinfo {author} {\bibfnamefont {A.}~\bibnamefont {Gopal}},
  \bibinfo {author} {\bibfnamefont {I.}~\bibnamefont {Watts}}, \bibinfo
  {author} {\bibfnamefont {F.~N.}\ \bibnamefont {Beg}}, \bibinfo {author}
  {\bibfnamefont {A.~E.}\ \bibnamefont {Dangor}}, \bibinfo {author}
  {\bibfnamefont {K.}~\bibnamefont {Krushelnick}}, \bibinfo {author}
  {\bibfnamefont {U.}~\bibnamefont {Wagner}}, \bibinfo {author} {\bibfnamefont
  {P.~A.}\ \bibnamefont {Norreys}}, \bibinfo {author} {\bibfnamefont {E.~L.}\
  \bibnamefont {Clark}}, \bibinfo {author} {\bibfnamefont {M.}~\bibnamefont
  {Zepf}}, \ and\ \bibinfo {author} {\bibfnamefont {R.~G.}\ \bibnamefont
  {Evans}},\ }\href {\doibase 10.1063/1.1469027} {\bibfield  {journal}
  {\bibinfo  {journal} {Physics of Plasmas}\ }\textbf {\bibinfo {volume} {9}},\
  \bibinfo {pages} {2244} (\bibinfo {year} {2002})},\ \Eprint
  {http://arxiv.org/abs/https://doi.org/10.1063/1.1469027}
  {https://doi.org/10.1063/1.1469027} \BibitemShut {NoStop}%
\bibitem [{\citenamefont {Aymar}, \citenamefont {Barabaschi},\ and\
  \citenamefont {Shimomura}(2002)}]{AymarPPCF2002}%
  \BibitemOpen
  \bibfield  {author} {\bibinfo {author} {\bibfnamefont {R.}~\bibnamefont
  {Aymar}}, \bibinfo {author} {\bibfnamefont {P.}~\bibnamefont {Barabaschi}}, \
  and\ \bibinfo {author} {\bibfnamefont {Y.}~\bibnamefont {Shimomura}},\ }\href
  {\doibase 10.1088/0741-3335/44/5/304} {\bibfield  {journal} {\bibinfo
  {journal} {Plasma Physics and Controlled Fusion}\ }\textbf {\bibinfo {volume}
  {44}},\ \bibinfo {pages} {519} (\bibinfo {year} {2002})}\BibitemShut
  {NoStop}%
\bibitem [{\citenamefont {{Khurana}}\ \emph {et~al.}(2004)\citenamefont
  {{Khurana}}, \citenamefont {{Kivelson}}, \citenamefont {{Vasyliunas}},
  \citenamefont {{Krupp}}, \citenamefont {{Woch}}, \citenamefont {{Lagg}},
  \citenamefont {{Mauk}},\ and\ \citenamefont {{Kurth}}}]{2004jpsm.book..593K}%
  \BibitemOpen
  \bibfield  {author} {\bibinfo {author} {\bibfnamefont {K.~K.}\ \bibnamefont
  {{Khurana}}}, \bibinfo {author} {\bibfnamefont {M.~G.}\ \bibnamefont
  {{Kivelson}}}, \bibinfo {author} {\bibfnamefont {V.~M.}\ \bibnamefont
  {{Vasyliunas}}}, \bibinfo {author} {\bibfnamefont {N.}~\bibnamefont
  {{Krupp}}}, \bibinfo {author} {\bibfnamefont {J.}~\bibnamefont {{Woch}}},
  \bibinfo {author} {\bibfnamefont {A.}~\bibnamefont {{Lagg}}}, \bibinfo
  {author} {\bibfnamefont {B.~H.}\ \bibnamefont {{Mauk}}}, \ and\ \bibinfo
  {author} {\bibfnamefont {W.~S.}\ \bibnamefont {{Kurth}}},\ }\enquote
  {\bibinfo {title} {{The configuration of Jupiter's magnetosphere}},}\ in\
  \href@noop {} {\emph {\bibinfo {booktitle} {Jupiter. The Planet, Satellites
  and Magnetosphere}}},\ Vol.~\bibinfo {volume} {1},\ \bibinfo {editor} {edited
  by\ \bibinfo {editor} {\bibfnamefont {F.}~\bibnamefont {{Bagenal}}}, \bibinfo
  {editor} {\bibfnamefont {T.~E.}\ \bibnamefont {{Dowling}}}, \ and\ \bibinfo
  {editor} {\bibfnamefont {W.~B.}\ \bibnamefont {{McKinnon}}}}\ (\bibinfo
  {year} {2004})\ pp.\ \bibinfo {pages} {593--616}\BibitemShut {NoStop}%
\bibitem [{\citenamefont {Valyavin}\ \emph {et~al.}(2014)\citenamefont
  {Valyavin}, \citenamefont {Shulyak}, \citenamefont {Wade}, \citenamefont
  {Antonyuk}, \citenamefont {Zharikov}, \citenamefont {Galazutdinov},
  \citenamefont {Plachinda}, \citenamefont {Bagnulo}, \citenamefont {Machado},
  \citenamefont {Alvarez}, \citenamefont {Clark}, \citenamefont {Lopez},
  \citenamefont {Hiriart}, \citenamefont {Han}, \citenamefont {Jeon},
  \citenamefont {Zurita}, \citenamefont {Mujica}, \citenamefont {Burlakova},
  \citenamefont {Szeifert},\ and\ \citenamefont {Burenkov}}]{ValyavinN2014}%
  \BibitemOpen
  \bibfield  {author} {\bibinfo {author} {\bibfnamefont {G.}~\bibnamefont
  {Valyavin}}, \bibinfo {author} {\bibfnamefont {D.}~\bibnamefont {Shulyak}},
  \bibinfo {author} {\bibfnamefont {G.~A.}\ \bibnamefont {Wade}}, \bibinfo
  {author} {\bibfnamefont {K.}~\bibnamefont {Antonyuk}}, \bibinfo {author}
  {\bibfnamefont {S.~V.}\ \bibnamefont {Zharikov}}, \bibinfo {author}
  {\bibfnamefont {G.~A.}\ \bibnamefont {Galazutdinov}}, \bibinfo {author}
  {\bibfnamefont {S.}~\bibnamefont {Plachinda}}, \bibinfo {author}
  {\bibfnamefont {S.}~\bibnamefont {Bagnulo}}, \bibinfo {author} {\bibfnamefont
  {L.~F.}\ \bibnamefont {Machado}}, \bibinfo {author} {\bibfnamefont
  {M.}~\bibnamefont {Alvarez}}, \bibinfo {author} {\bibfnamefont {D.~M.}\
  \bibnamefont {Clark}}, \bibinfo {author} {\bibfnamefont {J.~M.}\ \bibnamefont
  {Lopez}}, \bibinfo {author} {\bibfnamefont {D.}~\bibnamefont {Hiriart}},
  \bibinfo {author} {\bibfnamefont {I.}~\bibnamefont {Han}}, \bibinfo {author}
  {\bibfnamefont {Y.-B.}\ \bibnamefont {Jeon}}, \bibinfo {author}
  {\bibfnamefont {C.}~\bibnamefont {Zurita}}, \bibinfo {author} {\bibfnamefont
  {R.}~\bibnamefont {Mujica}}, \bibinfo {author} {\bibfnamefont
  {T.}~\bibnamefont {Burlakova}}, \bibinfo {author} {\bibfnamefont
  {T.}~\bibnamefont {Szeifert}}, \ and\ \bibinfo {author} {\bibfnamefont
  {A.}~\bibnamefont {Burenkov}},\ }\href@noop {} {\bibfield  {journal}
  {\bibinfo  {journal} {Nature}\ }\textbf {\bibinfo {volume} {515}},\ \bibinfo
  {pages} {88} (\bibinfo {year} {2014})}\BibitemShut {NoStop}%
\bibitem [{\citenamefont {Harding}\ and\ \citenamefont
  {Lai}(2006)}]{HardingRPP2006}%
  \BibitemOpen
  \bibfield  {author} {\bibinfo {author} {\bibfnamefont {A.~K.}\ \bibnamefont
  {Harding}}\ and\ \bibinfo {author} {\bibfnamefont {D.}~\bibnamefont {Lai}},\
  }\href {\doibase 10.1088/0034-4885/69/9/r03} {\bibfield  {journal} {\bibinfo
  {journal} {Reports on Progress in Physics}\ }\textbf {\bibinfo {volume}
  {69}},\ \bibinfo {pages} {2631} (\bibinfo {year} {2006})}\BibitemShut
  {NoStop}%
\bibitem [{\citenamefont {Potekhin}, \citenamefont {Pons},\ and\ \citenamefont
  {Page}(2015)}]{PotekhinSSR2015}%
  \BibitemOpen
  \bibfield  {author} {\bibinfo {author} {\bibfnamefont {A.~Y.}\ \bibnamefont
  {Potekhin}}, \bibinfo {author} {\bibfnamefont {J.~A.}\ \bibnamefont {Pons}},
  \ and\ \bibinfo {author} {\bibfnamefont {D.}~\bibnamefont {Page}},\
  }\href@noop {} {\bibfield  {journal} {\bibinfo  {journal} {Space Science
  Reviews}\ }\textbf {\bibinfo {volume} {191}},\ \bibinfo {pages} {239}
  (\bibinfo {year} {2015})}\BibitemShut {NoStop}%
\bibitem [{\citenamefont {Ichimaru}(2004)}]{Ichimaru2004}%
  \BibitemOpen
  \bibfield  {author} {\bibinfo {author} {\bibfnamefont {S.}~\bibnamefont
  {Ichimaru}},\ }\href {https://books.google.com/books?id=qqm3AAAAIAAJ} {\emph
  {\bibinfo {title} {Statistical Plasma Physics, Vol. 1}}}\ (\bibinfo
  {publisher} {Westview Press, Boulder, CO},\ \bibinfo {year}
  {2004})\BibitemShut {NoStop}%
\bibitem [{\citenamefont {Krall}\ and\ \citenamefont
  {Book}(1969)}]{KrallPF1969}%
  \BibitemOpen
  \bibfield  {author} {\bibinfo {author} {\bibfnamefont {N.~A.}\ \bibnamefont
  {Krall}}\ and\ \bibinfo {author} {\bibfnamefont {D.~L.}\ \bibnamefont
  {Book}},\ }\href {\doibase 10.1063/1.1692487} {\bibfield  {journal} {\bibinfo
   {journal} {The Physics of Fluids}\ }\textbf {\bibinfo {volume} {12}},\
  \bibinfo {pages} {347} (\bibinfo {year} {1969})},\ \Eprint
  {http://arxiv.org/abs/https://aip.scitation.org/doi/pdf/10.1063/1.1692487}
  {https://aip.scitation.org/doi/pdf/10.1063/1.1692487} \BibitemShut {NoStop}%
\bibitem [{\citenamefont {Vidal}\ and\ \citenamefont
  {Baalrud}(2021)}]{VidalPOP2021}%
  \BibitemOpen
  \bibfield  {author} {\bibinfo {author} {\bibfnamefont {K.~R.}\ \bibnamefont
  {Vidal}}\ and\ \bibinfo {author} {\bibfnamefont {S.~D.}\ \bibnamefont
  {Baalrud}},\ }\href {\doibase 10.1063/5.0045078} {\bibfield  {journal}
  {\bibinfo  {journal} {Physics of Plasmas}\ }\textbf {\bibinfo {volume}
  {28}},\ \bibinfo {pages} {042103} (\bibinfo {year} {2021})},\ \Eprint
  {http://arxiv.org/abs/https://doi.org/10.1063/5.0045078}
  {https://doi.org/10.1063/5.0045078} \BibitemShut {NoStop}%
\bibitem [{\citenamefont {Hollmann}, \citenamefont {Anderegg},\ and\
  \citenamefont {Driscoll}(1999)}]{HollmannPRL1999}%
  \BibitemOpen
  \bibfield  {author} {\bibinfo {author} {\bibfnamefont {E.~M.}\ \bibnamefont
  {Hollmann}}, \bibinfo {author} {\bibfnamefont {F.}~\bibnamefont {Anderegg}},
  \ and\ \bibinfo {author} {\bibfnamefont {C.~F.}\ \bibnamefont {Driscoll}},\
  }\href {\doibase 10.1103/PhysRevLett.82.4839} {\bibfield  {journal} {\bibinfo
   {journal} {Phys. Rev. Lett.}\ }\textbf {\bibinfo {volume} {82}},\ \bibinfo
  {pages} {4839} (\bibinfo {year} {1999})}\BibitemShut {NoStop}%
\bibitem [{\citenamefont {Ott}, \citenamefont {Bonitz},\ and\ \citenamefont
  {Donk\'o}(2015)}]{OttPRE2015}%
  \BibitemOpen
  \bibfield  {author} {\bibinfo {author} {\bibfnamefont {T.}~\bibnamefont
  {Ott}}, \bibinfo {author} {\bibfnamefont {M.}~\bibnamefont {Bonitz}}, \ and\
  \bibinfo {author} {\bibfnamefont {Z.}~\bibnamefont {Donk\'o}},\ }\href
  {\doibase 10.1103/PhysRevE.92.063105} {\bibfield  {journal} {\bibinfo
  {journal} {Phys. Rev. E}\ }\textbf {\bibinfo {volume} {92}},\ \bibinfo
  {pages} {063105} (\bibinfo {year} {2015})}\BibitemShut {NoStop}%
\bibitem [{\citenamefont {Kriesel}\ and\ \citenamefont
  {Driscoll}(2001)}]{KrieselPRL2001}%
  \BibitemOpen
  \bibfield  {author} {\bibinfo {author} {\bibfnamefont {J.~M.}\ \bibnamefont
  {Kriesel}}\ and\ \bibinfo {author} {\bibfnamefont {C.~F.}\ \bibnamefont
  {Driscoll}},\ }\href {\doibase 10.1103/PhysRevLett.87.135003} {\bibfield
  {journal} {\bibinfo  {journal} {Phys. Rev. Lett.}\ }\textbf {\bibinfo
  {volume} {87}},\ \bibinfo {pages} {135003} (\bibinfo {year}
  {2001})}\BibitemShut {NoStop}%
\bibitem [{\citenamefont {Scheiner}\ and\ \citenamefont
  {Baalrud}(2020)}]{ScheinerPRE2020}%
  \BibitemOpen
  \bibfield  {author} {\bibinfo {author} {\bibfnamefont {B.}~\bibnamefont
  {Scheiner}}\ and\ \bibinfo {author} {\bibfnamefont {S.~D.}\ \bibnamefont
  {Baalrud}},\ }\href {\doibase 10.1103/PhysRevE.102.063202} {\bibfield
  {journal} {\bibinfo  {journal} {Phys. Rev. E}\ }\textbf {\bibinfo {volume}
  {102}},\ \bibinfo {pages} {063202} (\bibinfo {year} {2020})}\BibitemShut
  {NoStop}%
\bibitem [{\citenamefont {Ott}\ and\ \citenamefont
  {Bonitz}(2011)}]{OttPRL2011}%
  \BibitemOpen
  \bibfield  {author} {\bibinfo {author} {\bibfnamefont {T.}~\bibnamefont
  {Ott}}\ and\ \bibinfo {author} {\bibfnamefont {M.}~\bibnamefont {Bonitz}},\
  }\href@noop {} {\bibfield  {journal} {\bibinfo  {journal} {Phys. Rev. Lett.}\
  }\textbf {\bibinfo {volume} {107}} (\bibinfo {year} {2011})}\BibitemShut
  {NoStop}%
\bibitem [{\citenamefont {Beck}, \citenamefont {Fajans},\ and\ \citenamefont
  {Malmberg}(1996)}]{BeckPOP1996}%
  \BibitemOpen
  \bibfield  {author} {\bibinfo {author} {\bibfnamefont {B.~R.}\ \bibnamefont
  {Beck}}, \bibinfo {author} {\bibfnamefont {J.}~\bibnamefont {Fajans}}, \ and\
  \bibinfo {author} {\bibfnamefont {J.~H.}\ \bibnamefont {Malmberg}},\
  }\href@noop {} {\bibfield  {journal} {\bibinfo  {journal} {Phys. Plasmas}\
  }\textbf {\bibinfo {volume} {3}},\ \bibinfo {pages} {1250} (\bibinfo {year}
  {1996})}\BibitemShut {NoStop}%
\bibitem [{\citenamefont {Cavalier}\ \emph {et~al.}(2013)\citenamefont
  {Cavalier}, \citenamefont {Lemoine}, \citenamefont {Bonhomme}, \citenamefont
  {Tsikata}, \citenamefont {Honoré},\ and\ \citenamefont
  {Grésillon}}]{CavalierPOP2013}%
  \BibitemOpen
  \bibfield  {author} {\bibinfo {author} {\bibfnamefont {J.}~\bibnamefont
  {Cavalier}}, \bibinfo {author} {\bibfnamefont {N.}~\bibnamefont {Lemoine}},
  \bibinfo {author} {\bibfnamefont {G.}~\bibnamefont {Bonhomme}}, \bibinfo
  {author} {\bibfnamefont {S.}~\bibnamefont {Tsikata}}, \bibinfo {author}
  {\bibfnamefont {C.}~\bibnamefont {Honoré}}, \ and\ \bibinfo {author}
  {\bibfnamefont {D.}~\bibnamefont {Grésillon}},\ }\href {\doibase
  10.1063/1.4817743} {\bibfield  {journal} {\bibinfo  {journal} {Physics of
  Plasmas}\ }\textbf {\bibinfo {volume} {20}},\ \bibinfo {pages} {082107}
  (\bibinfo {year} {2013})},\ \Eprint
  {http://arxiv.org/abs/https://doi.org/10.1063/1.4817743}
  {https://doi.org/10.1063/1.4817743} \BibitemShut {NoStop}%
\end{thebibliography}%

%\begin{thebibliography}{99}

% Effective Potential Theory for Transport Coefficients across Coupling Regimes
%\bibitem{baal:13} S.\ D.\ Baalrud and J.\ Daligault, Phys.\ Rev.\ Lett.\ {\bf 110}, 235001 (2013); Phys.~Rev.~E {\bf 91}, 063107 (2015).

%\end{thebibliography}

\end{document}